\documentclass[times]{simauth}
\usepackage{hyperref}
\usepackage[table]{xcolor}
\definecolor{Gray}{gray}{0.85}
\usepackage[parfill]{parskip}
\usepackage{titlesec}
\usepackage[superscript,biblabel]{cite}
\usepackage{caption}
\titleformat{\subsubsection}
  {\normalfont\normalsize\itshape}{\thesubsubsection}{1em}{}
\titlespacing*{\subsubsection}{0pt}{3.25ex plus 1ex minus .2ex}{0ex plus .2ex}

\begin{document}
\runninghead{G.D. and B.C.}
\title{Bayesian monotonic errors-in-variables models with applications to pathogen susceptibility testing}
\author{Glen DePalma\affil{a}\corrauth,
Bruce A. Craig\affil{b}}
\address{<\affilnum{a}Department of Statistics, Purdue University, 250 N. University Street, West Lafayette, IN 47907, USA\\
\affilnum{b}Department of Statistics, Purdue University, 250 N. University Street, West Lafayette, IN 47907, USA}
\corraddr{E-mail: glen.depalma@gmail.com}
\begin{abstract}
Drug dilution (MIC) and disk diffusion (DIA) are the two most common antimicrobial susceptibility assays used by hospitals and clinics to determine an unknown pathogen's susceptibility to various antibiotics.  Since only one assay is commonly used, it is important that the two assays  give similar results.  Calibration of the DIA assay to the MIC assay is typically done using the error-rate bounded method, which selects DIA breakpoints that minimize the observed discrepancies between the two assays.  In 2000, Craig proposed a model-based approach that specifically models the measurement error and rounding processes of each assay, the underlying pathogen distribution, and the true monotonic relationship between the two assays.  The two assays are then calibrated by focusing on matching the probabilities of correct classification (susceptible, indeterminant, and resistant).  This approach results in greater precision and accuracy for estimating DIA breakpoints.  In this paper, we expand the flexibility of the model-based method by introducing a Bayesian four-parameter logistic model (extending Craig's original three-parameter model) as well as a Bayesian nonparametric spline model to describe the relationship between the two assays.   We propose two ways to handle spline knot selection, considering many equally-spaced knots but restricting overfitting via a random walk prior and treating the number and location of knots as additional unknown parameters.  We demonstrate the two approaches via a series of simulation studies and apply the methods to two real data sets.
\end{abstract}
\maketitle

\section{Introduction}
\label{sec:intro}

The minimum inhibitory concentration (MIC) assay determines the lowest concentration of an antimicrobial agent that completely, or nearly completely, inhibits the growth of a microorganism. In practice, this concentration is compared to agent-specific MIC breakpoints to determine if the agent is a feasible treatment. Although this determination is not always correct, the MIC assay is accepted as the gold-standard reference to which all other testing methods are compared. 

One of the most widely-used alternative methods is the disk diffusion assay. It provides the diameter (DIA) of growth inhibition (the clear zone) around an antimicrobial disk of known potency placed on agar that has been inoculated with the microorganism.  Unfortunately there is not a straightforward conversion from diameter to MIC so a second set of breakpoints (DIA breakpoints) are needed in order for this assay to determine if an agent is a feasible treatment. 

Traditionally, the error-rate bounded (ERB) method has been used to calibrate the DIA assay to the MIC assay.  This procedure involves determining DIA breakpoints that minimize the observed discrepancies between results generated from both assays over a wide range of  pathogen strains (or isolates).  While simple and intuitive, this approach is very sample dependent and lacks precision \cite{annis205,BA}. 

Instead of focusing on observed assay results, model-based methods have been proposed to accurately describe the inherent variation in the observed results: measurement error, rounding, and the monotonic relationship between the two assays.  In 2000, Craig considered a  three-parameter logistic model to describe the true relationship \cite{BA}.  While Craig's approach works very well when a logistic curve adequately approximates the true underlying relationship, it also can produce poor and biased results when the function does not fit the true relationship well.  In 2008, Qi proposed a two-stage nonparametric approach \cite{xiaoli}.  First, he modeled the underlying MIC distribution using nonparametric density estimation based on M-splines. Given this density, he then fit the MIC/DIA relationship using I-splines.  While reasonable, there is concern that information is lost via this two-stage approach because the  pathogen density provides information on where the true DIA fall along the fitted curve and vice versa. 

In this paper, we extend the previously proposed model-based approaches by allowing for increased flexibility in the functional relationship between MIC and DIA assay results.  First, we extend Craig's model to a more flexible four-parameter logistic model.  Second, we consider a nonparametric spline model similar to Qi's but take a Bayesian approach that allows for the simultaneous estimation of the pathogen distribution and the relationship between the two assays.   For the spline fitting, we consider two approaches to knot selection.  One approach is to consider the number and location of knots as additional parameters and use a  reversible jump Markov chain Monte Carlo (RJMCMC) algorithm for estimation.  The second approach is to consider many equally-spaced knots and control overfitting through the use of a random-walk prior.

Even though the previous model-based approaches have been shown to be superior to the ERB, they have not been used in practice mainly due to the lack of available software.  We therefore developed the free software package \texttt{dBETS} (\textbf{d}iffusion \textbf{B}reakpoint \textbf{E}stimation \textbf{T}esting \textbf{S}oftware) that implements the proposed models.  The software can be accessed through an online interface: https://dbets.shinyapps.io/dBETS/. Information on the use of this software is described elsewhere \cite{DePalma16}.

The rest of the paper is organized as follows.  In the methods section, we describe the details of the two proposed model-based approaches.  The Bayesian inference section describes the Bayesian computation algorithm and model parameter priors.  We then assess the performance of the two models through simulation studies and application to two real data sets.  We conclude with a brief discussion.

\section{Methods}

\begin{figure}[!ht]
\captionsetup{justification=raggedright,singlelinecheck=false}
	\centering
	\includegraphics[scale=.5]{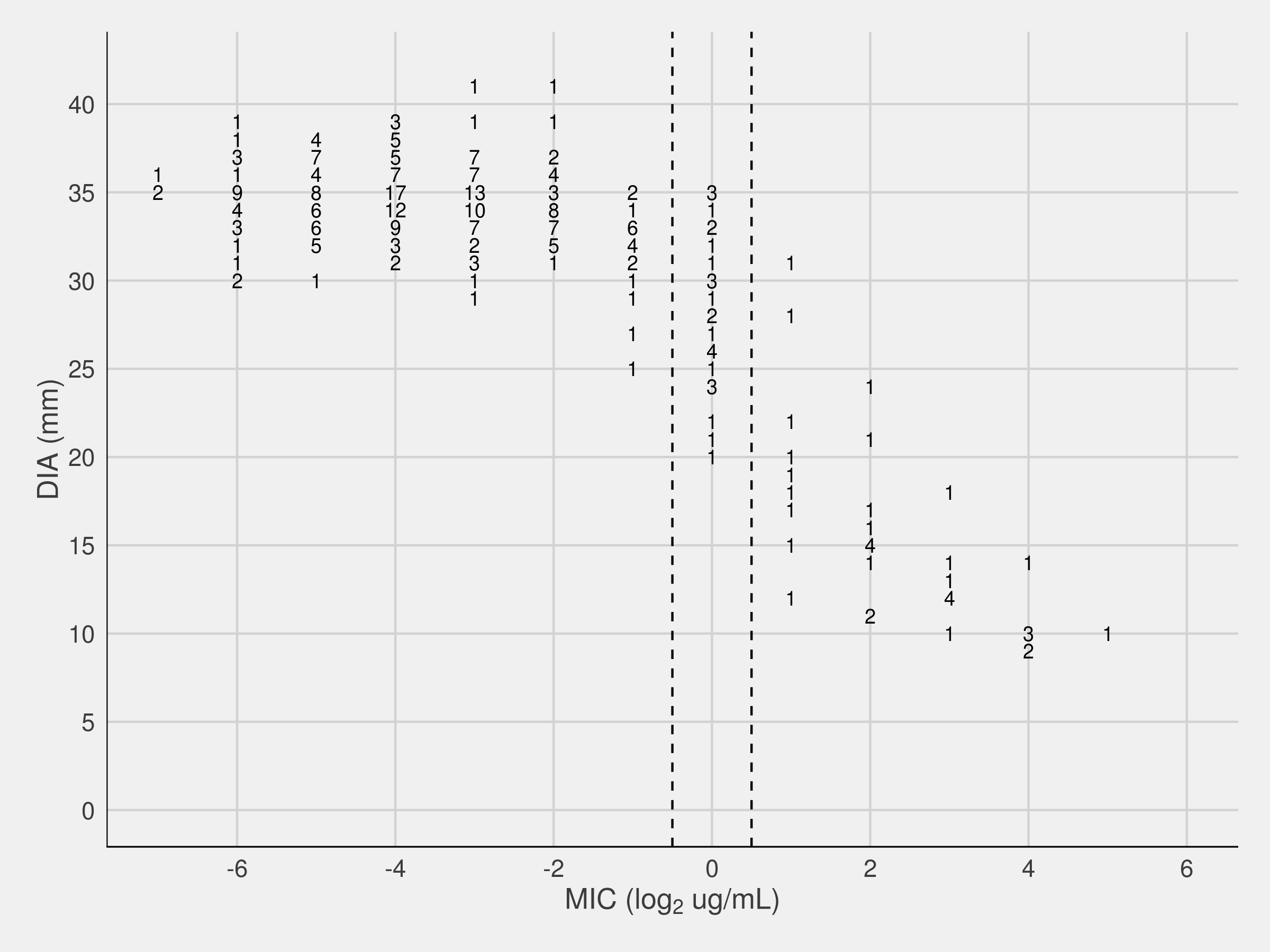}
	\caption{\normalsize Visual representation of a common DIA calibration study data set.  The numbers in the plot represent the number of isolates observed at that particular MIC/DIA combination.  The vertical dashed lines represent the MIC breakpoints.}
	\label{data1}
\end{figure}

Figure \ref{data1} is a scatterplot of 298 simulated DIA and MIC assay results.  It visually represents the  type of data set commonly used in a DIA calibration study.  The DIA results are reported in millimeters while the MIC results are reported in $\log_2$ concentrations.  Given MIC assay breakpoints (represented as dashed vertical lines), the goal of the study is to determine compatible DIA breakpoints.  These would be represented as horizontal dashed lines in the figure.  The area to the left of each set of breakpoints  is the susceptible region, the area in the middle indeterminant, and the area to the right is the resistant region. Compatible DIA breakpoints are those that in some way minimize the chance an isolate will fall into different classification regions based on assay.   

\newpage

The proposed and existing model-based procedures describe the observed pattern in the MIC/DIA scatterplot using a hierarchical model \cite{BA}.  These models account for the inherent variability of each assay (e.g., a 3-fold dilution range for the MIC assay) by estimating the true continuous (but unobservable) MIC and DIA values for each assay pathogen.  These true values are linked based on an assumed monotonically decreasing relationship between the true MIC and DIA \cite{murray09}.  

In other words, these hierarchical models describe the scatterplot in terms of three components:
\begin{enumerate}
\item The assay procedures (i.e., rounding and measurement error)
\item The drug/bug-specific relationship between the true MICs and DIAs
\item The underlying distribution of pathogens (or true MICs)
\end{enumerate}

\begin{figure}[!ht]
\captionsetup{justification=raggedright,singlelinecheck=false}
	\centering
	\includegraphics[scale=.5]{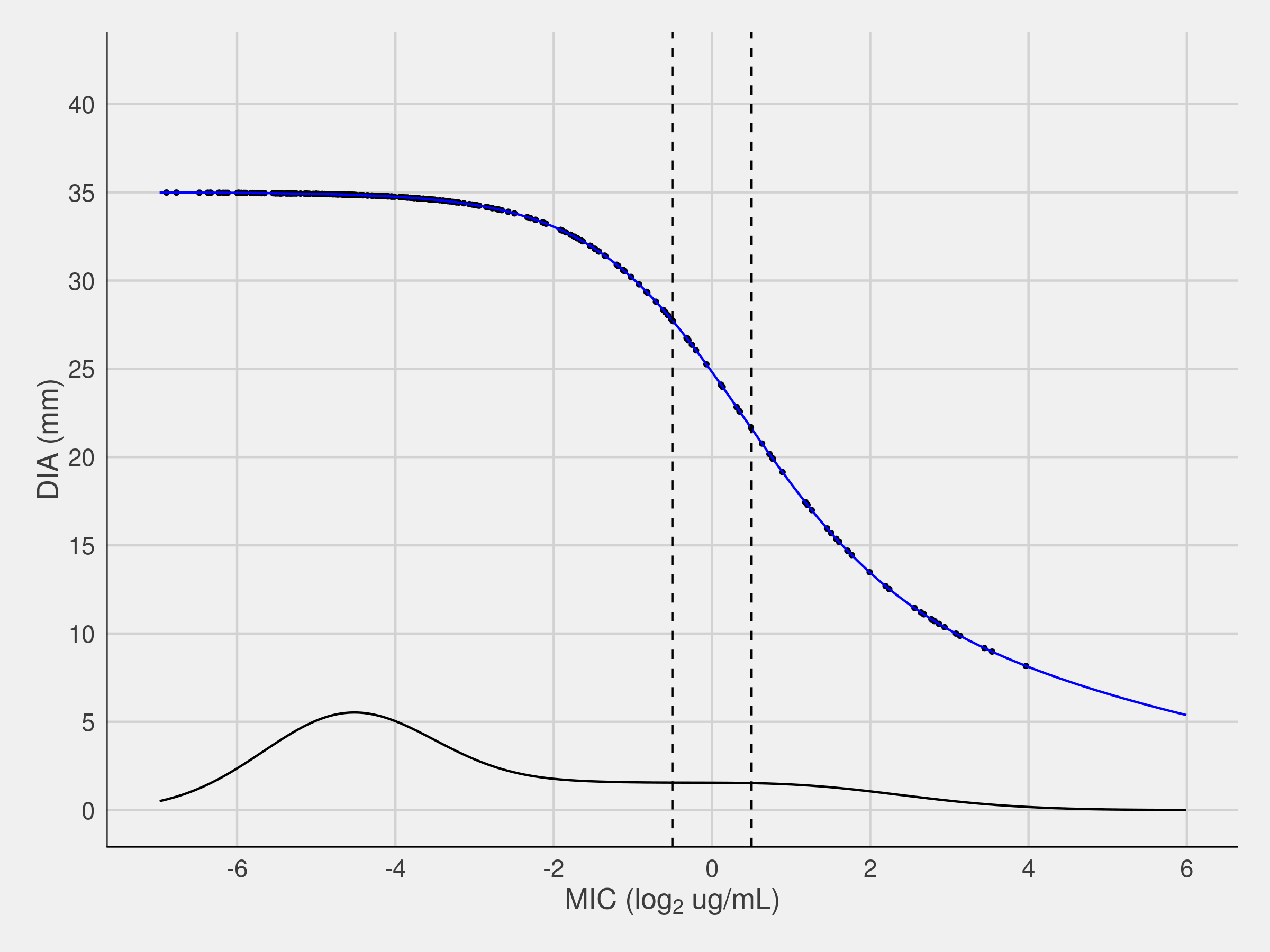}
	\caption{\normalsize A visual of the hierarchical model fit to the scatterplot in Figure 1.  The blue curve represents the monotonically decreasing true relationship between the true MIC and DIA values (Component 2).  Each black dot on this curve represents the true MIC for an observed MIC/DIA pair (Component 1).  The distribution of the true MICs is represented by the black curve (Component 3). }
\label{probModel}
\end{figure}

The first component links each observed MIC/DIA pair with an underlying true MIC value.  The second and third components describe the relationship between the true MIC and its corresponding true DIA.  Figure 
\ref{probModel} is a visual representation of Craig's logistic model fit to the data in Figure \ref{data1}.

\newpage

\subsection{Model Details}

With DIA breakpoint determination, one observes pairs of MIC and DIA assay results. These assay results are not perfect and therefore contain measurement error that needs to be accounted for when estimating the true underlying relationship.  In addition, the MIC assay rounds upwards to the nearest integer while the DIA assay rounds to the nearest integer \cite{annis205}.

\subsubsection{The assay procedures (i.e. rounding) and experimental variability}

For each pathogen $i$, we introduce a latent value $m_i$ to represent the corresponding true MIC value. The corresponding $d_i=g(m_i)$ represents the true latent diameter value. The joint distribution of observed MIC ($x_i$) and observed DIA ($y_i$) can then be expressed as:
\begin{align}
&x_i=\lceil m_i+\epsilon\rceil  \hspace{5mm} y_i=[g(m_i)+\delta] \nonumber \\
&\epsilon \sim N(0,\sigma_m^2) \hspace{6mm} \delta \sim N(0,\sigma_d^2)
\label{struModel}
\end{align}
where $\sigma_m$ and $\sigma_d$ represent the measurement error standard deviations for the MIC and DIA assays, respectively, the function $g()$ represents the true MIC and DIA relationship,  $\lceil . \rceil$ represents rounding up to the nearest integer (ceiling), and $[.]$ represents rounding to the nearest integer.

Past studies have shown that the measurement error distribution for the MIC and DIA are both Normal \cite{annis05,annis205} and we will assume this here.  Thus given $m_i$, the probability we observe $x_i$ is
\begin{align*}
\Phi(x_i;m_i,\sigma^2_m)-\Phi(x_i-1;m_i,\sigma^2_m)
\end{align*}

\noindent where $\Phi$ is the standard Normal CDF.  The difference in Normal CDFs accounts for the upwards rounding inherent in the assay.  Similarly, for that same isolate, $d_i=g(m_i)$ so the probability we observe $y_i$ is
\begin{align*}
\Phi(y_i+.5;d_i,\sigma^2_d)-\Phi(y_i-.5;d_i,\sigma^2_d)
\end{align*}

Because the pairs are independent, the complete likelihood of observed assay results given the vector of true MICs $\boldsymbol{m}$ and $g()$ is
\begin{align}
\prod_{i=1}^N & [\Phi(x_i;m_i,\sigma^2_m)-\Phi(x_i-1;m_i,\sigma^2_m)] \times
[\Phi(y_i+.5;g(m_i),\sigma^2_d)-\Phi(y_i-.5;g(m_i),\sigma^2_d) ]
\end{align}

\subsubsection{The relationship between the true MICs and DIAs}

We propose two models:

\begin{enumerate}
\item The true DIA and MIC values follow a four-parameter logistic curve.  This is a generalization of the three-parameter logistic model described in Craig 2000 \cite{BA}.  The four-parameter model is more flexible in that the shape of the curve does not have to be symmetric.
\item The true DIA and MIC values follow a nonparametric curve based on I-splines.  This method, inspired by Qi (2008), is more flexible and data-driven.
\end{enumerate}

\underline{Model 1}

The four-parameter logistic curve can be written as:

\begin{align}
g(m_i)=&\beta_1 \frac{w_i \exp(-\beta_3  
 (\beta_2-m_i))+(1-w_i)  \exp(-\beta_4  (\beta_2-m_i))} 
 {1+w_i  \exp(-\beta_3  (\beta_2-m_i))+(1-w_i)
 \exp(-\beta_4  (\beta_2-m_i))}
\end{align}
where
\begin{equation*}
w_i = 1/(1+\exp(-\beta^* (\beta_2-m_i))) \text{ and }   \beta^* = (2 \beta_3 \beta_4)/(\beta_3+\beta_4) 
\end{equation*}

This curve allows for asymmetry around the inflection point $\beta_2$ and reduces to the three-parameter model when $\beta_3=\beta_4$. It is a weighted average of two logistic curves with rates $\beta_3$ and $\beta_4$ and weights $w_i$ and $1-w_i$.  This parametrization is sometimes referred to as a five-parameter logistic model in the literature, because $\beta^*$ can be considered an additional parameter \cite{gottschalk05}.  In addition to the coefficients all being positive, the formulation of $\beta^*$ here guarantees a monotonically decreasing function as it can be written as the average of the reciprocals of $\beta_3$ and $\beta_4$ \cite{log5}.  

\underline{Model 2}

The spline model uses I-splines to describe the underlying relationship between the MIC and DIA assay.  I-splines are the CDF of M-splines, which have the property of a probability density function.  Because of this, I-splines are monotonically decreasing (or increasing) bases and a linear combination of them using only positive coefficients will ensure a monotonically decreasing (or increasing) function.  Ramsay (1988) provides elegant recursive algorithms for computing M-splines and I-splines \cite{Ramsay88spline}.

\begin{figure}[!ht]
\captionsetup{justification=raggedright,singlelinecheck=false}
	\centering
	\includegraphics[scale=.45]{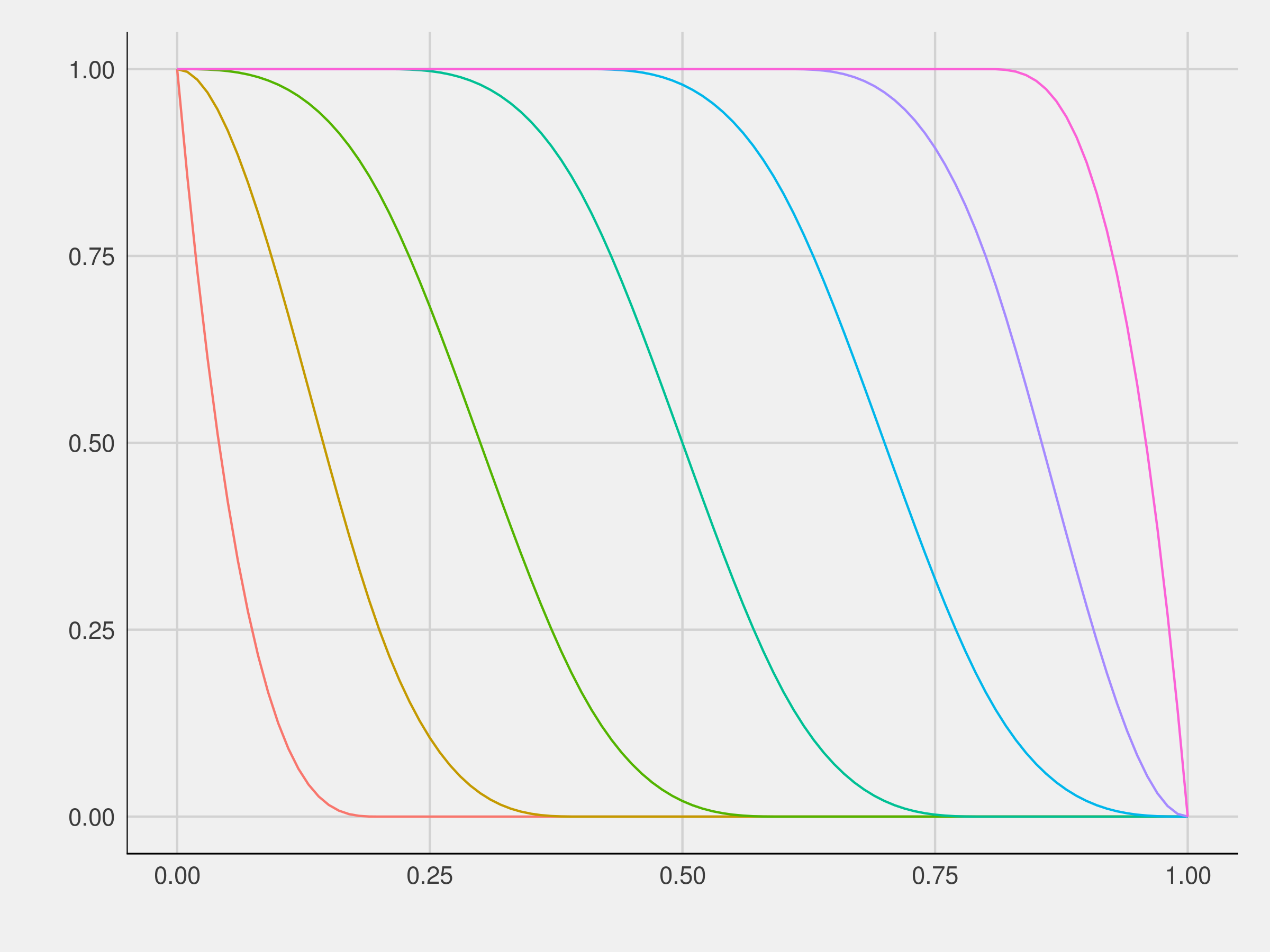}
	\caption{\normalsize Cubic I-splines ($1-I_i$) with knots at 0, .2, .4, .6, .8 and 1.  Monotonicity is enforced by the shape of the splines, provided the spline coefficients are nonnegative.}
	\label{ISplines}
\end{figure}

Given a knot sequence and bases ($\boldsymbol{I}$), the true DIA assay results are connected to the true MIC assay results by

\begin{equation}
d_i=g(m_i)=\sum_{j=1}^{j=B} \beta_j I_j(m_i)
\end{equation}
where $\boldsymbol{\beta}$ are the unknown spline coefficients and $B$ is the number of bases.  These bases are shown visually in Figure \ref{ISplines} for one knot sequence.   

\newpage

The choice of knot sequence is a key component to any spline procedure. If knots are not carefully selected, the fit can suffer. Spline methods will typically overfit (underfit) the data if too many (few) knots are selected. In addition, the knots need to be placed in optimal positions for the fit to be accurate. Common knot selection approaches are based on either fit statistics or cross validation \cite{spiriti08}. To find this set via fit statistics, one can place many equally-spaced knots and then find the knot sequence that minimizes a goodness of fit statistic, such as AIC or BIC. For cross-validation, a subset of data is fit to each specified knot sequence and the knot sequence that minimizes some sort of fit criterion, such as the mean squared error, is selected. This is the approach Qi (2008) took.  We describe our approaches in Section 3.1 because they introduce additional model parameters.

\subsubsection{The underlying distribution of pathogens (or MICs)}

We use a Dirichlet Process Mixture of Normals (DPMN) to model the underlying  distribution of pathogen true MICs $f(m)$ \cite{ghosh03}.  Intuitively the Dirichlet process (DP) can be thought of as a Bayesian mixture model consisting of $K$ components;
\begin{eqnarray*}
m_i|z_i,\boldsymbol{\mu},\boldsymbol{\sigma} \sim \mathcal{N}(\mu_{z_i},\sigma_{z_i})\\
z_i|\boldsymbol{p} \sim Multinomial(\boldsymbol{p}) \\
\boldsymbol{p}|\alpha \sim Dirichlet(\alpha/K,\dots,\alpha/K) \\
(\boldsymbol{\mu},\boldsymbol{\sigma}) \sim G_0
\label{DP}
\end{eqnarray*}
where $z_i$ is the group indicator, the vectors $\boldsymbol{\mu}$ and $\boldsymbol{\sigma}$ are the $K$ Normal means and standard deviations, respectively, the vector $\boldsymbol{p}$ is the mixing proportion, $\alpha$ is the concentration of the Dirichlet prior, and $G_0$ is the prior distribution of the component parameters.

Skipping the details, letting $K$ approach infinity leads to the Dirichlet Process Mixture (DPM), where the probabilities of an $m_i$ belonging to a cluster or forming a new cluster are proportional to the number in the current cluster and the $\alpha$ parameter \cite{escobar}.  This leads to rewriting the DPM as

\begin{eqnarray*}
m_i|\theta_i \sim \mathcal{N}(m_i|\mu_i,\sigma_i^2) \\
\theta_i=(\mu_i,\sigma_i) \sim G \\
G \sim DP(\alpha,G_0) \\
\end{eqnarray*}

where $G_0$ is known as the base distribution of the DP.  It specifies the prior distribution over $\mu$ and $\sigma$.  This rewriting circumvents having to specify the number of of components in the mixture model.  Neal (2000) provides various estimation methods for the Dirichlet Mixture Process \cite{Neal}.

\subsection{Determination of DIA Breakpoints}

Given a true MIC value and the hierarchical model that describes the distribution of the observed MIC results, the probability that an observed MIC will fall in any of the three classification regions can be computed. Traditionally, these regions have been defined by the MIC assay breakpoints.  Given the MIC assay properties, we make a distinction between assay MIC breakpoints and true MIC breakpoints.  Due to the upwards rounding of the MIC assay, we shift down the true MIC breakpoints by $0.5$ \cite{annis205}.  Thus if the assay MIC breakpoints are $-1$ and $1$ the true MIC breakpoints are $-1.5$ and $0.5$.  The assay MIC breakpoints are referred to as $M_L$ and $M_U$ while the true MIC breakpoints are referred to as $M_L^*$ and $M_U^*$.  This addition is a modification to the procedure proposed by Craig.

Similarly, given a set of DIA breakpoints, $D_U$ and $D_L$, and the true underlying relationship $g()$, the probability of an observed DIA falling in each region can also be calculated.  The model-based approaches focus on the probability of correct classification.  These probabilities, for given MIC $m$, are 

\begin{eqnarray}
\resizebox{.8\hsize}{!}{$
p_{{\mathrm MIC}}(m) = \left\{\begin{array}{ll}
{\mathrm Pr}(x \leq M_{L}) = \Phi\left(\frac{M_{L}-m}{\sigma_{m}}\right) &
m \leq M_L^* \\
{\mathrm Pr}(M_{L} < x < M_{U}) = \Phi\left(\frac{M_{U}-1-m}{\sigma_{m}}\right) -
\Phi\left(\frac{M_{L}-m}{\sigma_{m}}\right) & M_L^* < m < M_U^* \\
{\mathrm Pr}(x \geq M_{U}) = 1-\Phi\left(\frac{M_{U}-1-m}{\sigma_{m}}\right) &
m \geq M_U^* 
\end{array}\right.
$}
\end{eqnarray}
\begin{eqnarray}
\resizebox{.8\hsize}{!}{$
p_{{\mathrm DIA}}(m) = \left\{\begin{array}{ll}
{\mathrm Pr}(y \geq D_{U}) = 1-\Phi\left(\frac{D_{U}-.5-g(m)}{\sigma_{d}}\right) &
m \leq M_L^* \\
{\mathrm Pr}(D_{L} < y < D_{U}) = \Phi\left(\frac{D_{U}-.5-g(m)}{\sigma_{d}}\right) -
\Phi\left(\frac{D_{L}+.5-g(m)}{\sigma_{d}}\right) & M_L^* < m < M_U^* \\
{\mathrm Pr}(y \leq D_{L}) = \Phi\left(\frac{D_{L}+.5+g(m)}{\sigma_{d}}\right) &
m \geq M_U^*
\end{array}\right.
$}
\end{eqnarray}

The optimal DIA breakpoints are the set that minimizes a weighted loss function.  Our loss function is the accumulated  squared difference in the probability of correct classification when the DIA assay performs worse than the MIC assay. The underlying MIC density is used as the weighting function.  This can be expressed:
\begin{equation}
L =
\int_{-\infty}^{\infty}{\mathrm{min}\left(0,p_{\mathrm{DIA}}(g(u))-p_{\mathrm{MIC}}(u)\right)^{2}w(u) \, du}
\label{lossFunc}
\end{equation}

If $p_{\mathrm{DIA}}(g(u))$ is always greater than $p_{\mathrm{MIC}}(u)$, meaning the DIA assay outperforms the MIC assay, the DIA breakpoints will be set as wide as possible.  Thus, we are no longer treating the MIC assay as the gold standard. 

\section{Model Estimation}

With any Bayesian analysis, prior distributions are required on the parameters \cite{BIDA}.  When possible, we have chosen relatively noninformative priors for each of the parameters.  Bayesian analysis produces posterior distributions of the parameters which in turn produce a posterior distribution of the set of DIA breakpoints. 

To estimate the model parameters from a scatterplot, we take a Bayesian approach using Markov chain Monte Carlo (MCMC).  We now detail the estimation process of the nonparametric spline model.  The estimation process for the four-parameter logistic model is very similar, with the only difference being in the parameters describing the true MIC/DIA relationship.  That process is similar to the one described in Craig (2000) for the three-parameter logistic function.   

\subsection{Model Parameters}

In this analysis, the true MIC values are unknown and treated as additional parameters.  Since we are not dealing with a fixed data set, this complicates the knot selection process as selection techniques based on fit statistics do not suffice.  We propose two solutions: (1)  consider a set of equally-spaced interior knots across the range of MIC values and constrain the coefficients via a random walk prior to avoid overfitting \cite{BIDA}, and (2) consider the number and locations of interior knots as additional unknown parameters \cite{DiMatteo01}. Because we are considering cubic splines, both approaches have three knots fixed at the the lowest observed $\mathrm{MIC}-0.5$ and the largest observed $\mathrm{MIC}+0.5$.  These two approaches introduce additional model parameters to be estimated.

The following is a list of unknown model parameters that must be estimated from the scatterplot of data:
\begin{itemize}
\item[1. ] $\boldsymbol{m}$ - the true MIC values
\item[2. ] $\boldsymbol{\beta}$ - the I-spline coefficients
\item[3a. ] $\lambda$ -  the smoothing parameter for a fixed number of equally spaced knots
\item[3b. ]  $k$ and $\boldsymbol{t}$  - the number of interior knots and their locations
\item[4. ] $f(\boldsymbol{m})$ -the underlying pathogen MIC distribution
\end{itemize}

\subsubsection{Bayesian Computation}
\label{sec:Bayesian}
When considering a fixed set of equally spaced knots, we're interested in the posterior $\pi(\boldsymbol{\beta},f(\boldsymbol{m}),\lambda|\boldsymbol{x},\boldsymbol{y},\sigma_m,\sigma_d)$.  To make this computation easier we consider the individual true MICs as additional parameters.  By doing so, the posterior can be broken down and is proportional to
\begin{equation}
\pi(\boldsymbol{x},\boldsymbol{y}, \sigma_m, \sigma_d|\boldsymbol{\beta},\boldsymbol{m}) \times 
\pi(\boldsymbol{m}|f(\boldsymbol{m})) \times
\pi(f(\boldsymbol{m})|DP(\alpha,G_0)) \times
\pi(DP(\alpha,G_0),\boldsymbol{\beta},\lambda)
\end{equation}

The model estimation process is detailed in Appendix \ref{Appen: Estimation}.

\subsubsection{Priors}
When considering many knots at equally spaced intervals, we constrain the coefficients via a random walk prior \cite{BIDA}. The log of each  coefficient has a Normal prior with mean equal to the log of the current coefficient estimate with $\lambda$ serving as a smoothness parameter.  For $\beta_1$ we consider the non-informative $\mathcal{LN}(0,100)$ prior and place interior knots every $0.5$ units.  For the other coefficients,   
\begin{align}
\log{\beta_{j+1}} | \beta_1,\dots,\beta_{j} \sim \mathcal{N}(\log{\beta_{j}}, \lambda)
\label{randomWalk}
\end{align}
This prior forces each spline coefficient to be positive and relatively close to the previous spline coefficient to avoid overfitting.

When considering the number and location of interior knots as unknown parameters, we add a process to add, delete, or move them within our Markov chain.  This is performed using Reversible Jump MCMC (RJMCMC) \cite{Green}.  Current proposals have been based on a fixed set of points \cite{DiMatteo01}, however in our case, the true assay values are updated during each step of the MCMC process.  Therefore RJMCMC updates are based on the least-squares coefficient estimates with the thinking that the current coefficients will be close to the least-squares coefficients.  The Jacobian reduces to 1 during each step of the process (See Appendix: \ref{Appen: Estimation}).  The prior for $\boldsymbol{\beta}$ is Normal to improve mixing, however the resulting  fit is checked to be monotonically decreasing during the update step.

In summary, the priors for the two nonparametric approaches are as follows:

\begin{itemize}
\item Spline Model 1: Equally spaced knots
\begin{enumerate}
\item $\pi(\beta_1) \sim \mathcal{LN}(0,100)$
\item $\pi(\log{\beta_j}|\beta_{j-1},\lambda) \sim \mathcal{N}(\log{\beta_{j-1}},\lambda) \; j=2, \dots, B$
\item $\pi(\lambda) \sim Uniform(0,2)$
\end{enumerate}
\item Spline Model 2: Treating the number and position of interior knots as unknown parameters
\begin{enumerate}
\item $\pi(\boldsymbol{\beta}) \sim \mathcal{N}(0,100)$
\begin{itemize}
\item $\boldsymbol{\beta}$ is a vector of length $k+3=B$
\end{itemize}
\item $k \sim TPOI(3)$ where $TPOI(3)$ is a truncated Poisson distribution (no probability is assigned at $0$) with $\lambda=3$.  
\begin{itemize}
\item This results in a mean number of interior knots of $3.15$ with a $95\%$ confidence interval of $1$ to $7$.
\end{itemize}
\item $t_j \sim Uniform(t_{j-1},t_{j+1}) \; j=1, \dots, k$
\begin{itemize}
\item The locations $t_0$ and $t_{k+1}$ are the exterior knot locations.  
\end{itemize}
\end{enumerate}
\item Four-parameter Logistic Model
\begin{enumerate}
\item $\pi(\boldsymbol{\beta}) \sim \mathcal{LN}(0,100)$
\end{enumerate}
\item Priors for the Dirichlet Process Mixture of Normals - Both Models
\begin{enumerate}
\item $\alpha \sim Uniform(.2,2)$
\item $G_0$ on $\theta=(\mu,\sigma^2)$
\begin{itemize}
\item $\mu \sim \mathcal{N}(0,100)$
\item $\sigma^2 \sim InvGamma(.01.01)$
\end{itemize}
\end{enumerate}
\end{itemize}

\section{Results}

We perform a simulation study to assess the performance of the logistic four-parameter model and the nonparametric spline model.  For the spline model, we use equally spaced knots and constrain the coefficients via the random walk prior (Model 1).  Performance is similar for the RJMCMC model (Model 2).

\subsection{Simulation Details}
\label{Sim Details}

DIA breakpoint determination can be affected by many factors, such as the underlying pathogen density, the true underlying relationship between the MIC and DIA, the degree of measurement error in each assay, the location of MIC breakpoints, and the number of pathogens.  For this first study, we examine four different MIC/DIA relationships and MIC density combinations summarized in Table \ref{scenarios}.  

For Scenario 1, the true linear relationship is $\boldsymbol{d}=30-2\boldsymbol{m}$.  For Scenario 2, the four-parameter logistic coefficients are (35,1.17, 0.1, 1.2).  For Scenario 3, the I-spline knots are set at $(-3,0,1)$ and the coefficients set at $(1,1,20,1,20,1)$.  For Scenario 4, the I-spline knots are set at $(-4,-2,0,2,4)$ and the coefficients set at $(1,10,1,25,1,1,10,1)$.   These scenarios were selected to cover a wide range of possible relationships and are similar to what we have seen experimentally.

\begin{table}[h!]
\captionsetup{justification=raggedright,singlelinecheck=false}
\begin{center}
\begin{tabular}{ l l c c c}
\hline
& \textbf{MIC-DIA Relationship} &  \multicolumn{3}{c}{\textbf{MIC Density Parameters}} \\
&& \multicolumn{1}{c}{$\boldsymbol{\mu}$}  & \multicolumn{1}{c}{$\boldsymbol{\sigma}$}& \multicolumn{1}{c}{$\boldsymbol{\pi}$} \\
\hline
  Scenario 1 & Linear  & (-6, 3) & (2, 0.7) &(0.8, 0.2) \\
\rowcolor{Gray}   Scenario 2 & 4-Parameter Logistic & (-4.6,-2,1) & (0.6,0.2,0.2) & (1.1,1.5,1.5)\\
  Scenario 3 & I-Spline & (-3 ,0, 3) & (1, 1, 1) & (0.5, 0.3, 0.2) \\
\rowcolor{Gray}   Scenario 4 & I-Spline & (-3, 0, 3) & (2, 2, 2) & (1/3, 1/3, 1/3) \\
\bottomrule
\end{tabular}
\end{center}
\caption{Description of the simulation scenarios run in this analysis.}
\label{scenarios}
\end{table}

For each scenario, the measurement error standard deviations used for the MIC and DIA assays were set at $0.707$ and $2.121$, respectively, based on an abundance of quality control data  \cite{annis05,annis205}.  We used two sets of MIC breakpoints per scenario to evaluate DIA breakpoint performance.  The number of pathogens was set at $1000$ and we consider $200$ scatterplots.  The generation of each scatterplot involved:
\begin{enumerate}
\item Randomly sampling 1000 MIC isolates from the specified MIC density.
\item Determining the true DIA using the known MIC/DIA relationship.
\item Generating the observed assay pair by adding independent measurement errors to each true value and then rounding appropriately.
\end{enumerate}

We compare the performance of the two methods by looking at how close the estimated breakpoints are to the true breakpoints and the sum of squared errors (SSE) defined as:
\[
SSE=\int_{c_1}^{c_2} (\hat{h}(c)-h(c))^2 \, dc
\]
where  $h(.)$ is the true underlying density or functional relationship, and $\hat{h}(.)$ is the estimated density or relationship. The integration region $c_1$ to $c_2$ was taken to be the endpoints shown in Figures \ref{fitLog} and \ref{fitSpline} and the integral was approximated across a grid of $1000$ equally spaced points.

For each of the 200 scatterplots, model estimation took roughly 3 minutes using a combination of R and C.  The logistic model was about 30 seconds faster due to the estimation of fewer parameters.  The number of iterations was set at 12,000 with a burn-in of 6000.  For the data sets we looked at, this was adequate for the parameters to converge around their target values.

\subsection{Simulation Results}
\label{sec:SimResults}

To summarize the performance of each simulation, we use the median posterior estimate of $f()$ and $g()$.  During each iteration, $f()$ and $g()$ are evaluated along a MIC grid of $1000$ points.  We calculate the median of $f()$ and $g()$ for each MIC grid point and call this our median posterior estimate.

\begin{figure}[h!]
\captionsetup{justification=raggedright,singlelinecheck=false}
\begin{centering}
\begin{tabular}{cc}
  \includegraphics[scale=.35]{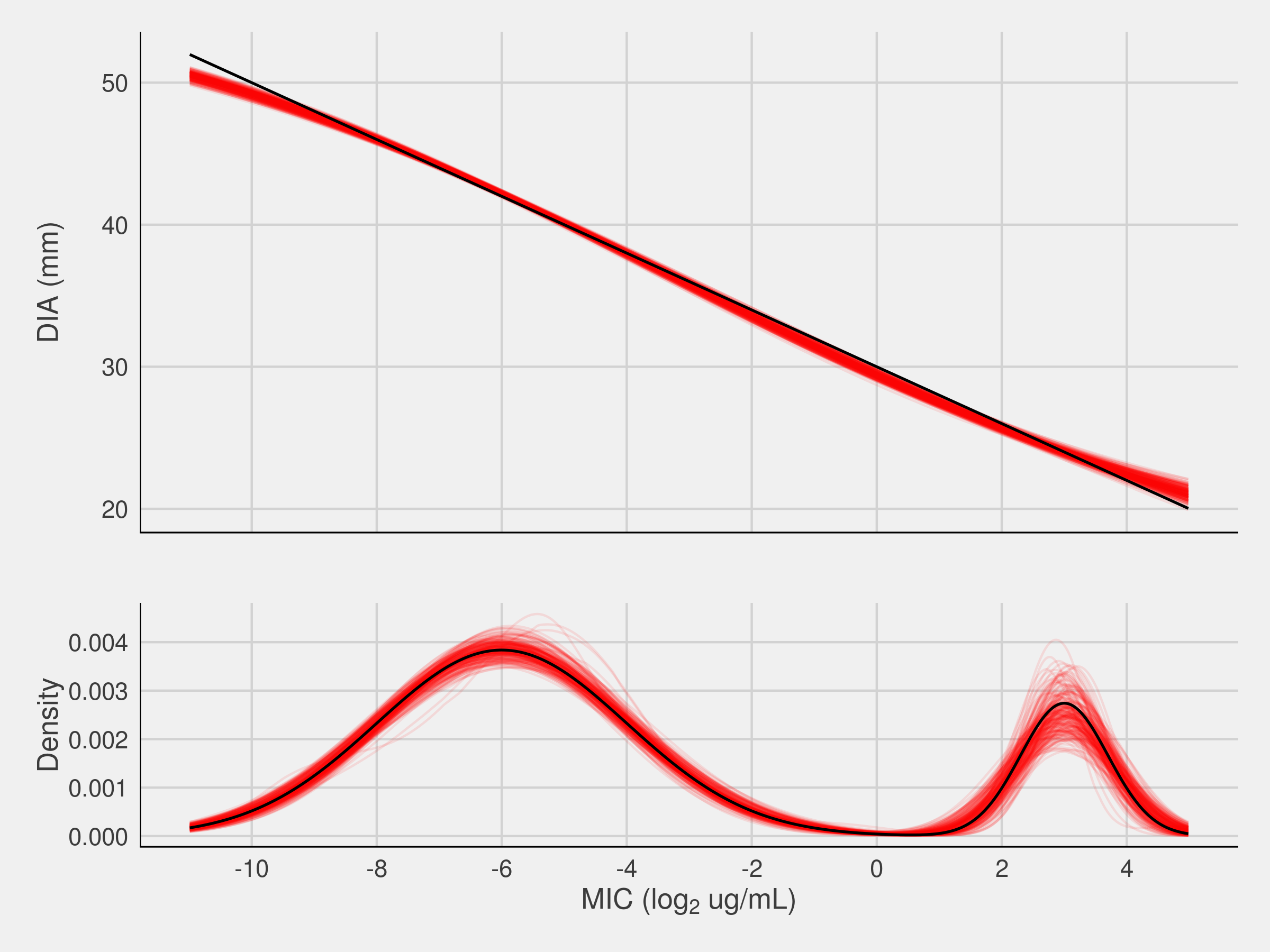} &   \includegraphics[scale=.35]{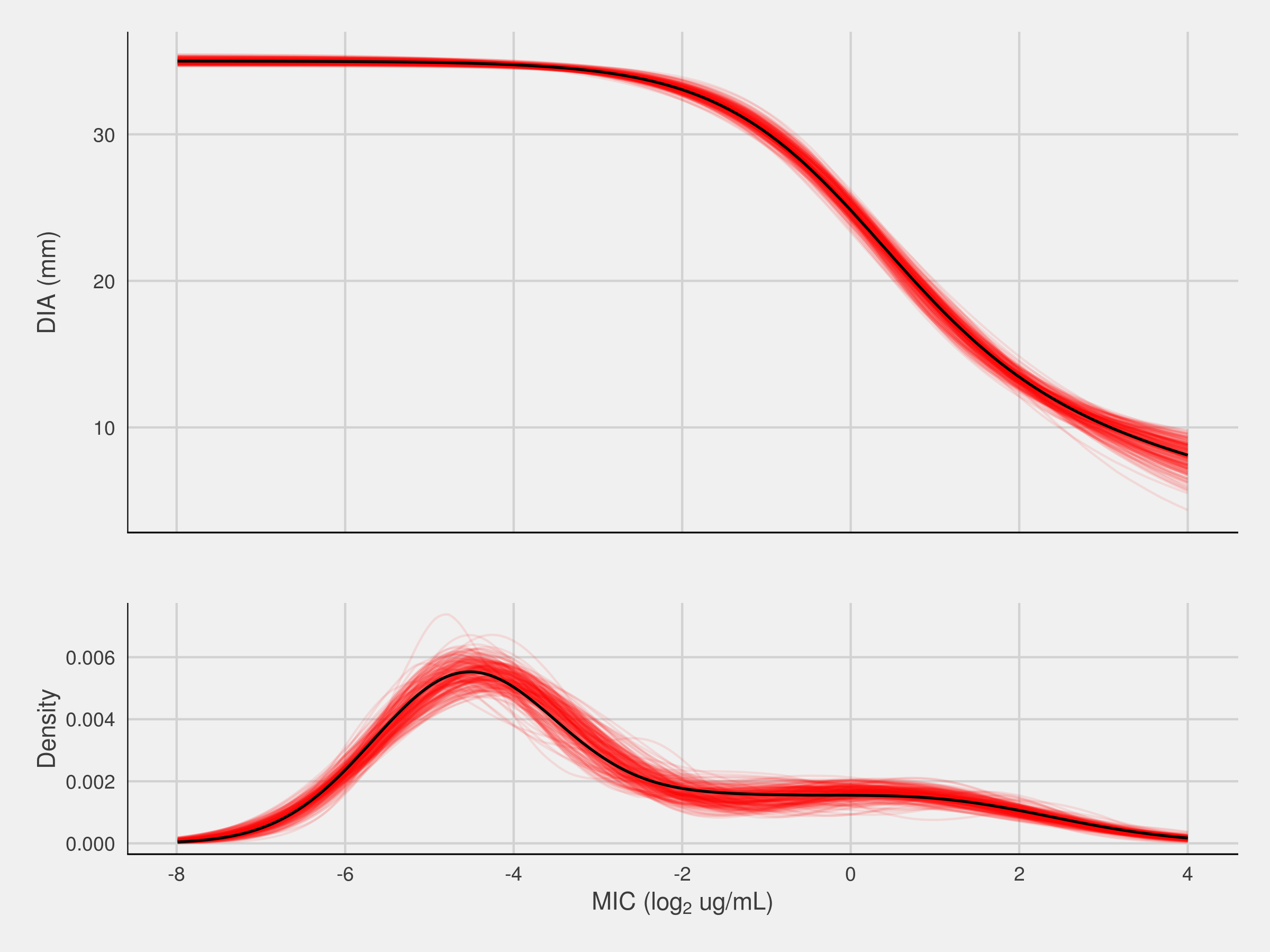} \\
(a) Scenario 1 & (b) Scenario 2 \\[6pt]
 \includegraphics[scale=.35]{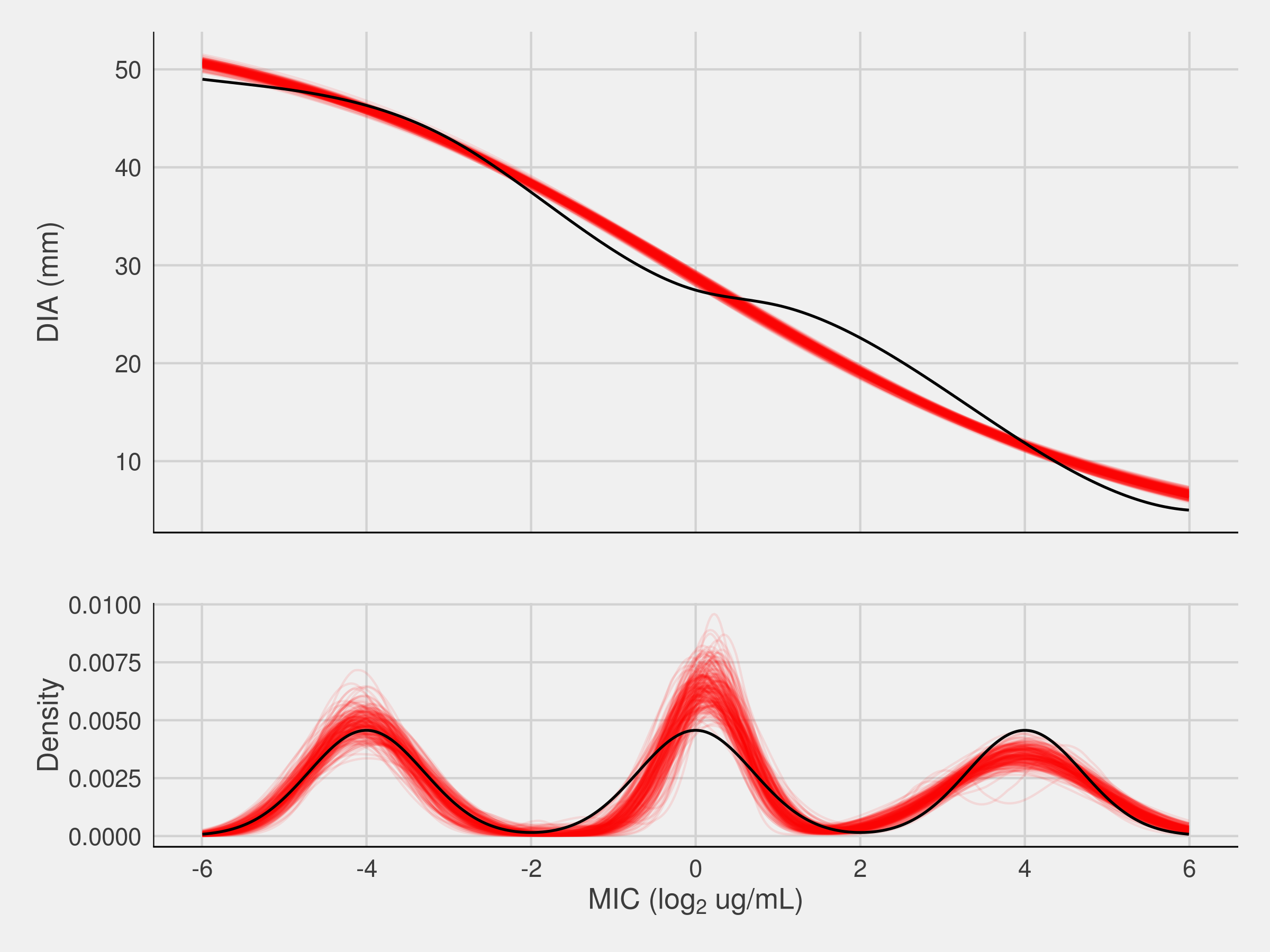} &   \includegraphics[scale=.35]{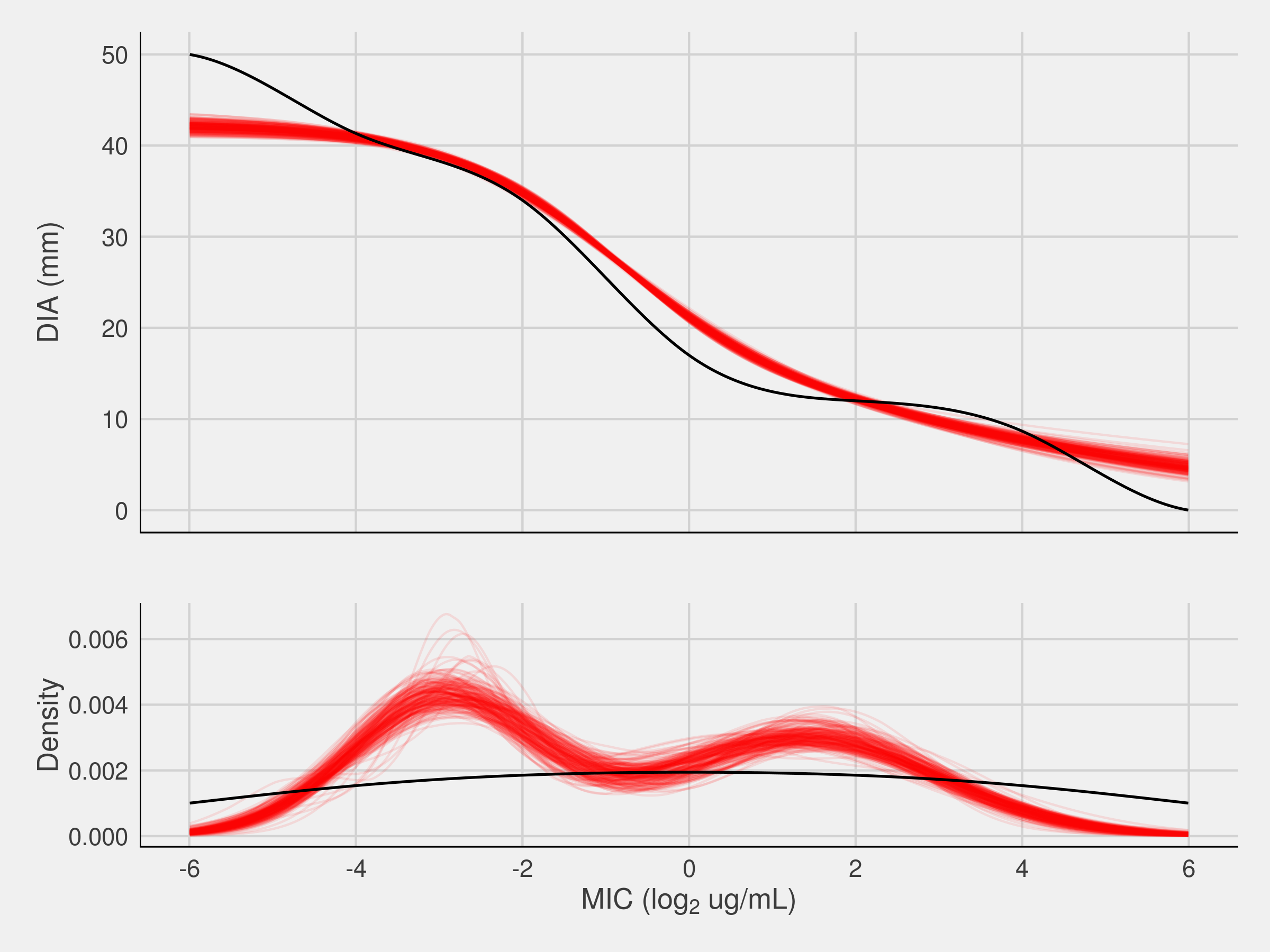} \\
(c) Scenario 3 & (d) Scenario 4 \\[6pt]
\end{tabular}
\caption{\normalsize Logistic regression results for the four simulation scenarios.  The black curve is the truth and the red curves are the median posterior estimates.}
\label{fitLog}
\end{centering}
\end{figure}

\begin{figure}[h!]
\captionsetup{justification=raggedright,singlelinecheck=false}
\begin{centering}
\begin{tabular}{cc}
  \includegraphics[scale=.35]{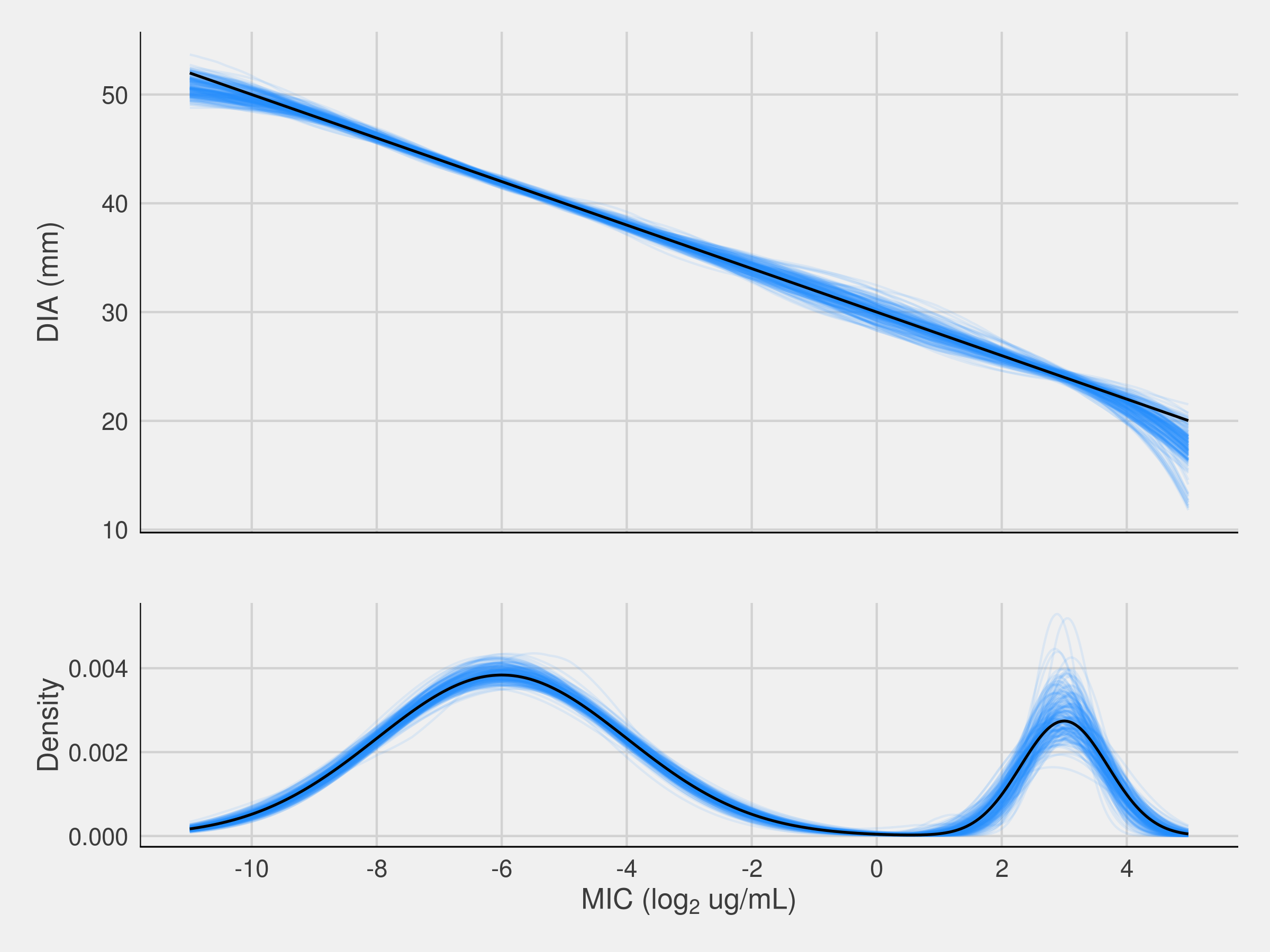} &   \includegraphics[scale=.35]{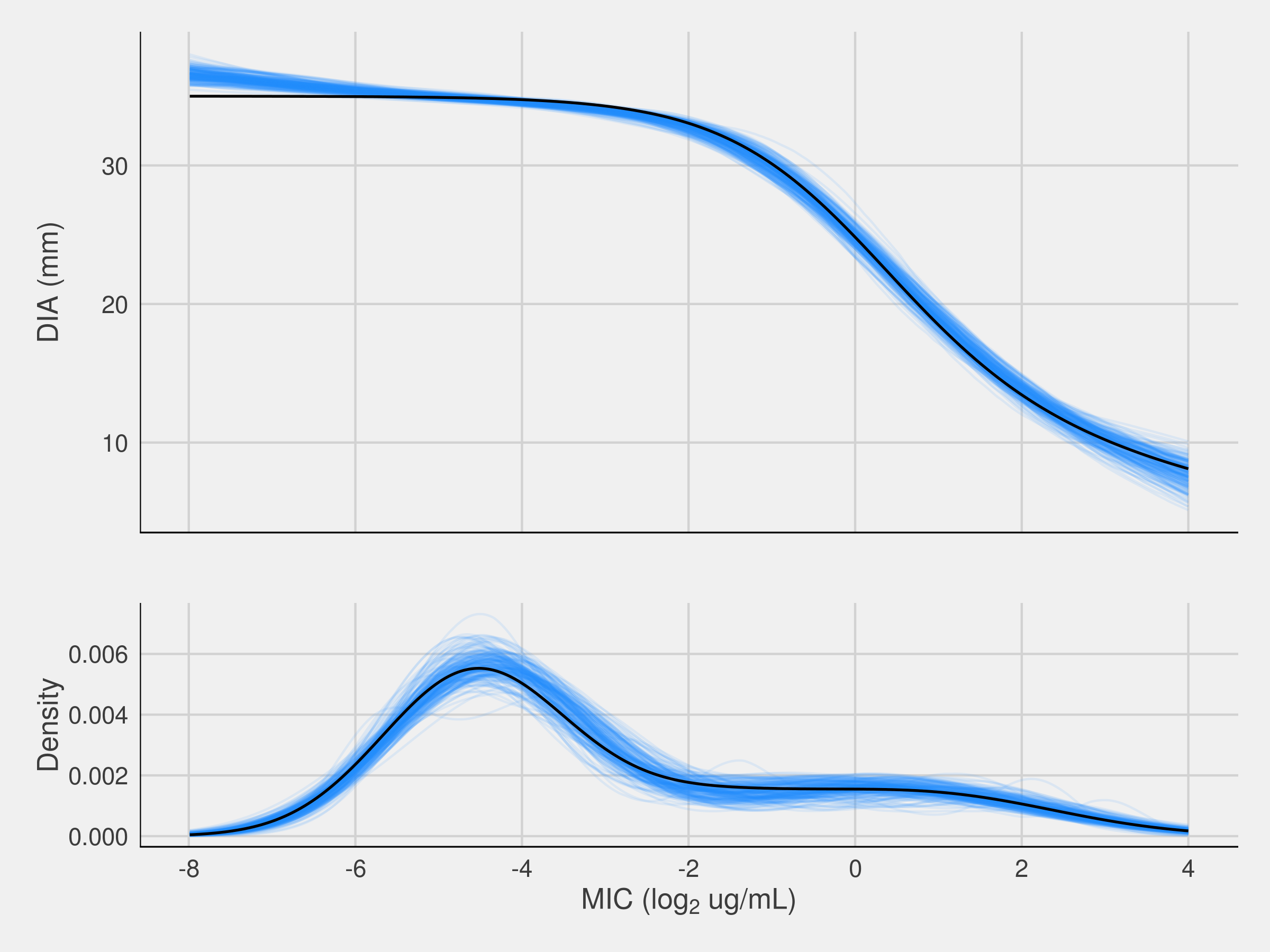} \\
(a) Scenario 1 & (b) Scenario 2 \\[6pt]
 \includegraphics[scale=.35]{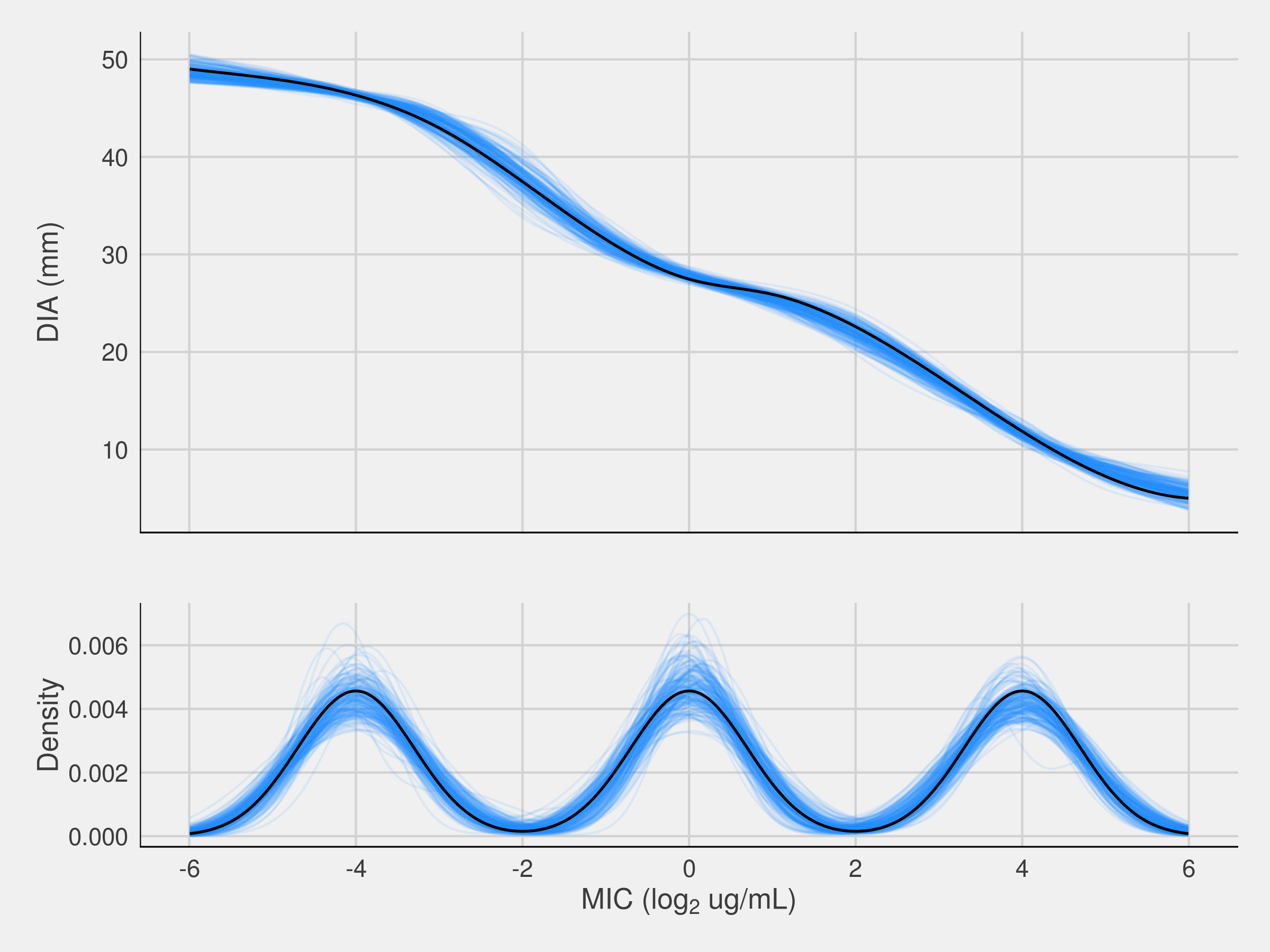} &   \includegraphics[scale=.35]{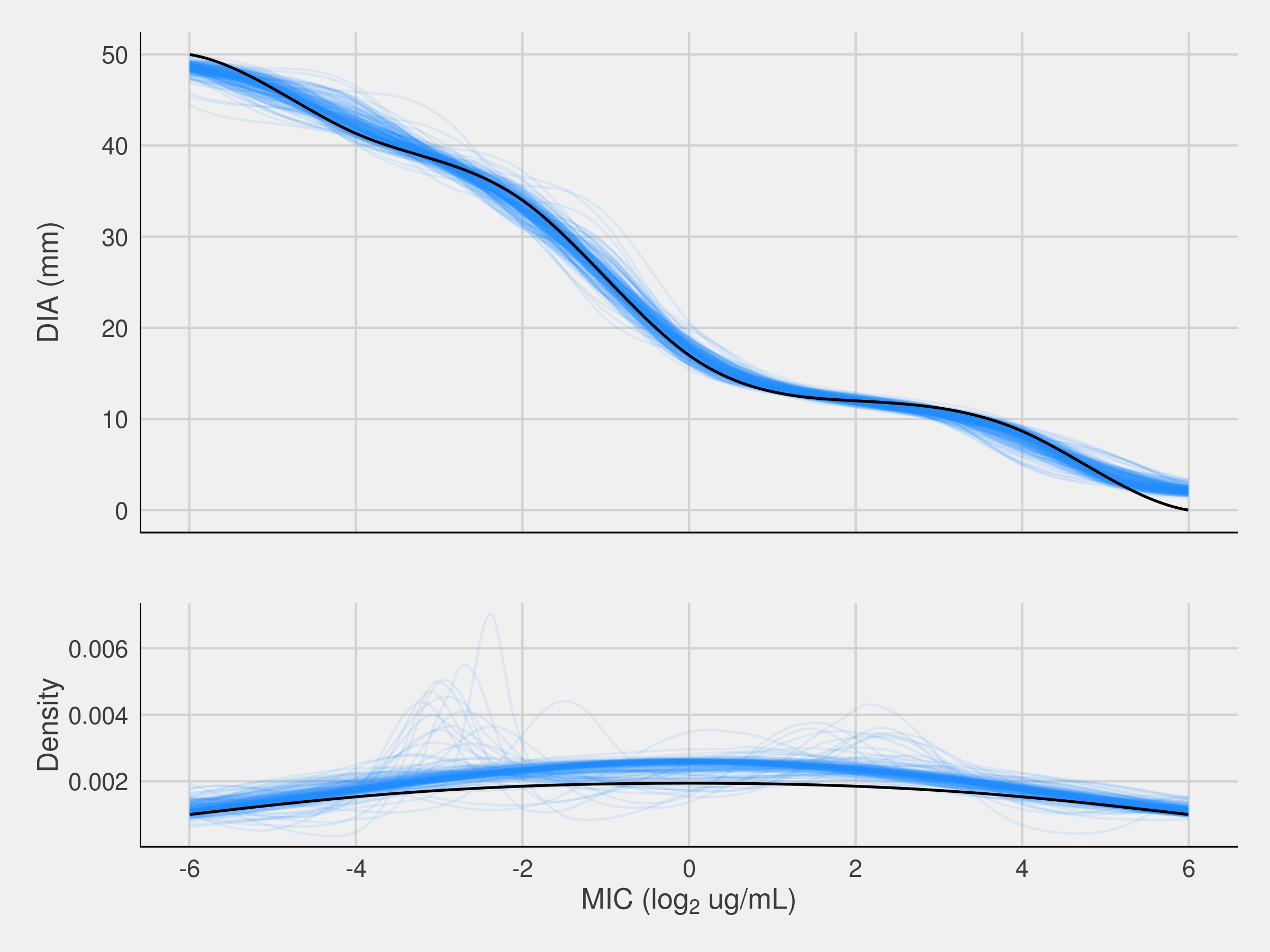} \\
(c) Scenario 3 & (d) Scenario 4 \\[6pt]
\end{tabular}
\caption{\normalsize Spline regression results for the four simulation scenarios.  The black curve is the truth and the blue curves are the median posterior estimates from 200 runs.}
\label{fitSpline}
\end{centering}
\end{figure}

Figures \ref{fitLog} and \ref{fitSpline} show the model fits for each simulation scenario for the logistic and spline models.  Table \ref{fitStat} summarizes the fit statistics for each model and simulation scenario.

\newpage

\begin{table}[h!]
\captionsetup{justification=raggedright,singlelinecheck=false}
\begin{center}
\begin{tabular}{ l c c | c c}
\hline
& \multicolumn{2}{c}{ SSE MIC-DIA} & \multicolumn{2}{c}{ SSE MIC Density} \\
& \multicolumn{1}{c}{\textbf{Logistic}} & \multicolumn{1}{c}{\textbf{Spline}} & \multicolumn{1}{c}{\textbf{Logistic}} & \multicolumn{1}{c}{\textbf{Spline}}  \\
\hline
  Scenario 1 & 129 -	113 -	57 & 156 -	105 -	164 & 0.21 - 0.16 - 0.18 & 0.12	- 0.10 -	0.10\\
\rowcolor{Gray}   Scenario 2 & 51 -	39- 34 & 62	- 44 - 55 & 0.26 - 0.21 - 0.18 & 0.44 - 0.36 - 0.33\\
  Scenario 3 & 1509	- 1504 - 135 & 171	- 148	- 110 & 2.66 - 2.41 - 1.23 & 0.47 - 0.45 - 0.21 \\
 \rowcolor{Gray}  Scenario 4 & 1954 - 1938 - 126 & 310 - 223 - 293 & 3.97 - 3.78 - 0.98 & 0.34 - 0.20 - 0.37\\
\bottomrule
\end{tabular}
\end{center}
\caption{Fit statistics for the simulation study.  Reporting Mean-Median-Standard Deviation.  The MIC density statistics were multiplied by 1000 for easier reading.}
\label{fitStat}
\end{table}

Performance is comparable in Scenario 1, with a slight edge to the spline model.  As expected, since the true relationship is a four-parameter logistic curve, the logistic model outperforms the spline model in Scenario 2 but not by much.  When the true model deviates from the parametric curves, the spline model performs much better as in Scenarios 3 and 4.  

Table \ref{brkptPerf} shows the percentage of times the estimated DIA breakpoints are close to the true DIA breakpoints for the two sets of assay MIC breakpoints.  Since the Bayesian analysis produces a posterior distribution of possible sets of DIA breakpoints, we select the maximum a posteriori probability set (MAP) for this analysis.  Essentially, this is the set of DIA breakpoints with the highest posterior probability.

\newpage

\begin{table}[h!]
\captionsetup{justification=raggedright,singlelinecheck=false}
\begin{center}
\begin{tabular}{l c c c c c}
\hline
& MIC Brkpts & \multicolumn{2}{c}{Exact Match DIA} & \multicolumn{2}{c}{Within 1 DIA} \\
& & \textit{Logistic} & \textit{Spline} & \textit{Logistic} & \textit{Spline} \\
\hline
  Scenario 1 & -6, -4 & 100 & 99 & 100 & 100 \\
 & 0, 2 & 82 & 86 & 100 & 100 \\
\hline
\rowcolor{Gray}   Scenario 2 & -2, 0 & 89 & 74 & 100 & 100 \\
\rowcolor{Gray}   & 0, 2 & 71 & 52 & 100 & 96 \\
\hline
  Scenario 3 & -1, 1 & 3 & 44 & 91 & 97  \\
& 1, 3 & 0 & 40 & 18 & 93  \\
\hline
\rowcolor{Gray}   Scenario 4 & -1, 1 & 0 & 37 & 0 & 95  \\
\rowcolor{Gray}    & 0, 2 & 0 & 38 & 0 & 92  \\
\bottomrule
\end{tabular}
\end{center}
\caption{Breakpoint performance for the simulation study.  The table shows the percentage of times the  MAP posterior DIA breakpoints match, or are within one of the true DIA breakpoints.}
\label{brkptPerf}
\end{table}

For Scenario 1, the performance between the two models in terms of correct DIA breakpoints is about equal.  Both models find the correct DIA breakpoints when the MIC breakpoints are $(-6, -4)$  more frequently.  This is most likely due to the fact that the breakpoint set $(0, 2)$  is in a relatively sparse data area.  For Scenario 2, the logistic model only outperforms the spline model in terms of finding the exact breakpoints. As in Scenario 1, both models find the exact breakpoints more frequently for the MIC breakpoint set located in a more data dense area.  

In Scenarios 3 and 4, the spline model does a much better job of estimating the DIA breakpoints and the logistic model performs poorly.  This suggests that it is crucial to accurately estimate the true relationship.  Interestingly, there is little difference in performance of the spline model for the two sets of breakpoints.  In Scenario 3, the logistic model still performs better for the breakpoints in a more data dense region.  

We tested the robustness of the results by only using 500 isolates instead of 1000.  Results were very similar, however SSE increased slightly (and DIA breakpoint accuracy decreased slightly) for both models as expected.  

\subsection{Gaps in the Data}
\label{gaps}

Three of the four scenarios in the previous section had areas of sparse MIC data. Because DIA breakpoint estimation was poor when the true relationship is not accurately estimated and when the MIC breakpoints are in the sparse data region, we decided to investigate this further by simulating a larger region where there is an absence of data.  

For the first simulation (Scenario 1), $g()$ is a three-parameter logistic model with coefficients $49$, $1.17$, and $0.4$.  The second simulation (Scenario 2), $g()$ is the I-spline model in Scenario 3 of Table \ref{scenarios}.  The true MICs were drawn from a mixture of two Normals with means $(-5.5, 5.5)$ for the first scenario and $(-4,4)$ for the second scenario.  The standard deviations were $(1,1)$ for the first scenario and $(0.5,0.5)$ for the second scenario.  We once again generated 200 scatterplots but only used 500 isolates.    We evaluate performance based on the SSE between the estimated and true MIC/DIA fit and the breakpoint estimation.  Figure \ref{gaps} shows one scatterplot result for each $g()$ relationship.

\newpage
\begin{figure}[h!]
\captionsetup{justification=raggedright,singlelinecheck=false}
\begin{centering}
\begin{tabular}{cc}
  \includegraphics[scale=.3]{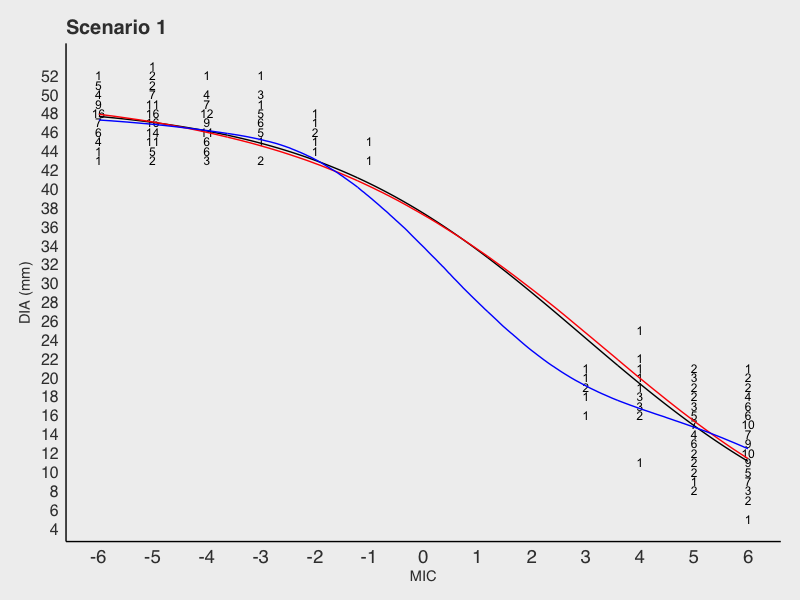} &   \includegraphics[scale=.3]{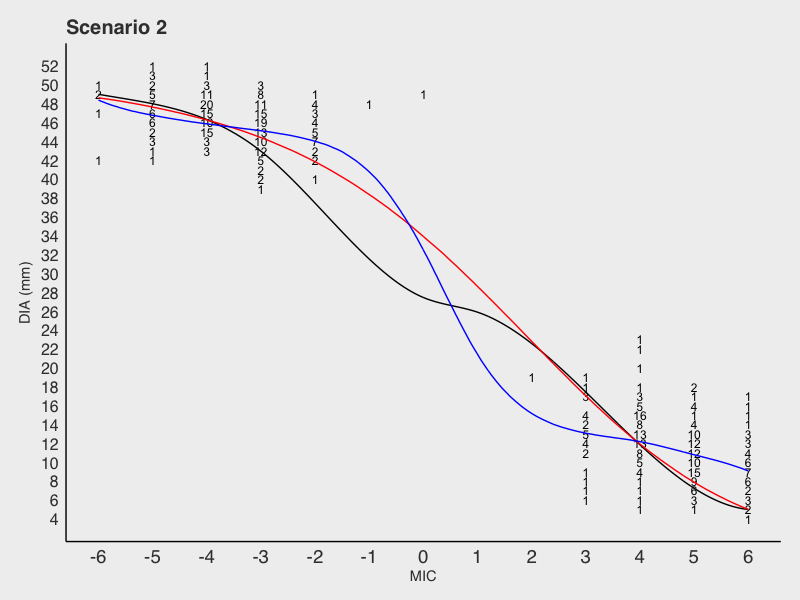} 
\end{tabular}
\caption{\normalsize Model fits for one simulation of the gap study for Scenario 1 (left) and Scenario 2 (right).  The black curve is the truth, the red curve is the estimated logistic fit, and the blue curve is the estimated spline fit.}
\label{gaps}
\end{centering}
\end{figure}

For Scenario 1, we see the logistic curve is able to estimate the true relationship quite well over the entire data region.  This is not too surprising given the relationship is logistic and the two data regions provide enough shape information.  The spline model does reasonably well fitting the data region on the left but struggles to fit the gap region and the data region on the right.  This also is not surprising as the spline weights nearby data more heavily in estimating the fit.  

Performance is not nearly as good in Scenario 2 for either model.  For the spline model, the general shape of the fit is similar to that in Scenario 1.  It is basically fitting each data region separately and connecting these two fits with a smoothed line.  For the logistic model, the fit is poor because the logistic cannot fit this shape well and the data regions do not supply enough shape information.  However, because the data regions are not that informative, the logistic model actually fits the extremes of the MIC data range better than it did previously (see Figure \ref{fitLog}, bottom left).      

These patterns were consistent across the 200 scatterplots.  Overall the logistic model had much lower SSEs than the spline model.  For Scenario 1, the mean SSE for the spline model was 19.3 times larger.  For Scenario 2, the mean SSE was 1.8 times larger.  

Ultimately, we are interested in how the absence of data affects breakpoint performance.  We used three sets of MIC breakpoints for each scenario: $(-5,-3)$, $(-1,1)$, and $(3,5)$.  DIA breakpoint performance is summarized in Table \ref{brkptPerfGap}.

\begin{table}[h!]
\captionsetup{justification=raggedright,singlelinecheck=false}
\begin{center}
\begin{tabular}{l c c c c c}
\hline
& MIC Brkpts & \multicolumn{2}{c}{Exact Match DIA} & \multicolumn{2}{c}{Within 1 DIA} \\
&& \textit{Logistic} & \textit{Spline} & \textit{Logistic} & \textit{Spline} \\
\hline
 & -5, -3 & 40 & 13 & 96 & 72  \\
  Scenario 1 & -1, 1 & 46 & 0 & 88 & 5 \\
 & 3, 5 & 71 & 0 & 100 & 0  \\
\hline
\rowcolor{Gray}  & -5, -3 & 60 & 9 & 94 & 18 \\
\rowcolor{Gray}   Scenario 2& -1, 1 & 0 & 0 & 0 & 0 \\
\rowcolor{Gray} & 3, 5 & 64 & 0 & 100 & 95 \\
\bottomrule
\end{tabular}
\end{center}
\caption{Breakpoint performance for the gap simulation study.  The table shows the percentage of times the  MAP posterior DIA breakpoints match, or are within one of the true DIA breakpoints.}
\label{brkptPerfGap}
\end{table}

These results reveal several interesting findings.  First, the logistic model performance is quite good and outperforms the spline model in every case.  Even in Scenario 2, when the true relationship is not logistic, the logistic model outperforms the spline model for each set of MIC breakpoints.  The logistic model also  performs relatively well in the missing data region for Scenario 1 but not for Scenario 2.  Second, the spline model struggles even when the the MIC breakpoints are in one of the data clouds.  This poor performance may be caused by the model basically fitting each data region separately and connecting the two with a smoothed line, and/or the general poor performance of nonparametric methods near the endpoints of the data region.

These results suggest that when there is an absence of data or the MIC breakpoints are located near the endpoints, the logistic model is preferred. However, one should be very careful when there is an absence of data.  If the true relationship differs from a logistic curve then the breakpoints will still be inaccurate.

\subsection{Selecting a Model}
\label{select}

In practice we recommend fitting both models to the data and performing the following steps.  First, examine the posterior distribution of breakpoints for each of the models.  There may be considerable overlap and the two models may essentially be in agreement.   

Second, look at the visual fits of the models to the data.  If the spline model is quite different than the logistic model this could be an indication that the logistic model is not appropriate.  However, if the MIC breakpoints are near the extremes of the data range, the logistic model is preferred.  If there are gaps in the data near the MIC breakpoints then proceed with caution.  The logistic model is preferred, however these breakpoints may still be poor if the logistic function is not a good approximation of the true relationship.  

Another case when the logistic model may be preferred is when some of the data points are censored. In practice, only a fixed set of concentrations are used in the MIC assay so if a true MIC is less than the lower concentration or greater than or equal to the upper concentration, the observed MIC may be censored.  The DIA test can also result in censored values if the diameter of the clear zone does not exceed the disk diameter or the clear zone gets too large.  Most of the time the censoring is not close to the MIC breakpoints and therefore is not a major concern in selecting a model.  If the MIC breakpoints are at the extremes of the MIC data range, the logistic model is preferred because the logistic model can predict true MIC values outside the data range and we can adjust the likelihood in Equation 2 to handle the censoring.   The nonparametric spline model should not be used to infer true MIC outside the data range.  As a result, for our spline calculations here, we ignore the censoring.

\subsection{Application to Real Data Sets}
\label{realData}

We demonstrate the proposed methods on two publicly available data sets from the  Clinical and Laboratory Standards Institute  (CLSI) summarized in Table \ref{tableDataSources}.  These are considered challenging data sets from which to estimate DIA breakpoints because of their features.

\begin{table}[ht!]
\captionsetup{justification=raggedright,singlelinecheck=false}
\centering
\begin{tabular}{l l l l r r r r r}
\hline
&&&&&\multicolumn{4}{c}{Breakpoints} \\
Organism & Agent & Abbreviation & Date & N & \multicolumn{2}{r}{MIC} & \multicolumn{2}{c}{DIA} \\ \hline
\textit{Enterobacteriaceae} & Ertapenem & ERT EB &  06/2011 & 948 & -1 & 1 & 18 & 20 \\
\rowcolor{Gray} \textit{Escherichia coli} & Pradofloxacin & PRA ECOL &  06/2013 & 312 & -2 & 1 & 19 & 24  \\
\bottomrule
\end{tabular}
\caption{Summaries of two real data sets. For MIC breakpoints (mg/L), the first value is the susceptible breakpoint and the second value is  the resistant breakpoint.  For DIA breakpoints (mm), the first value is the  resistant breakpoint and the second value is the  susceptible breakpoint.  The listed DIA breakpoints are what CLSI determined.}
\label{tableDataSources}
\end{table}

\begin{figure}[ht!]
\captionsetup{justification=raggedright,singlelinecheck=false}
\begin{centering}
\begin{tabular}{cc}
  \includegraphics[scale=.25]{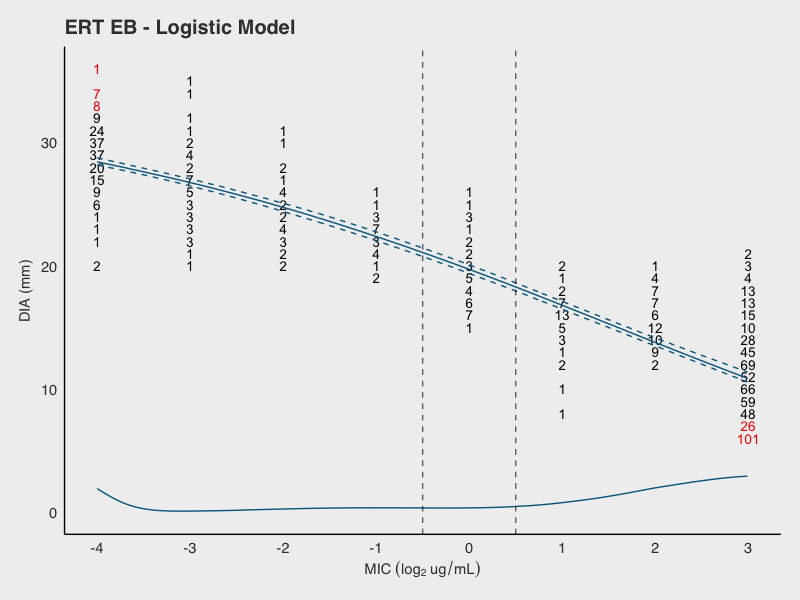} &   \includegraphics[scale=.25]{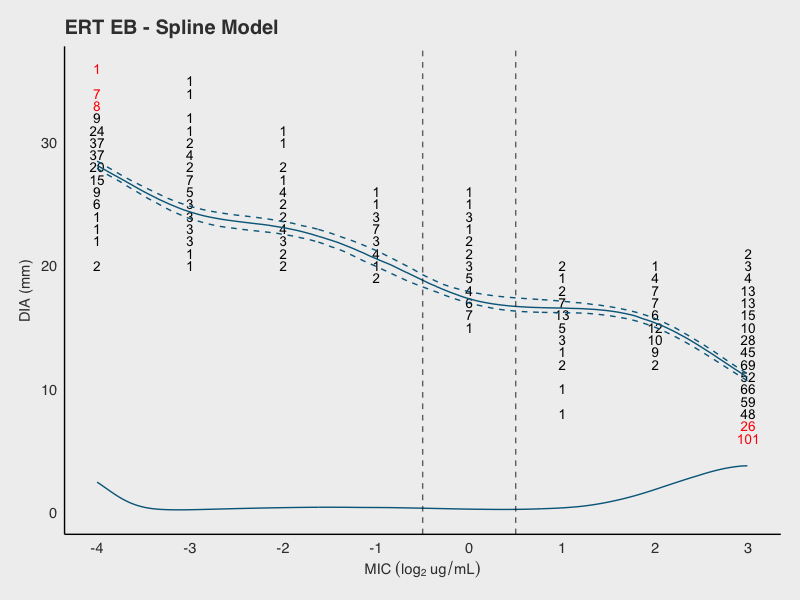} \\
 \includegraphics[scale=.25]{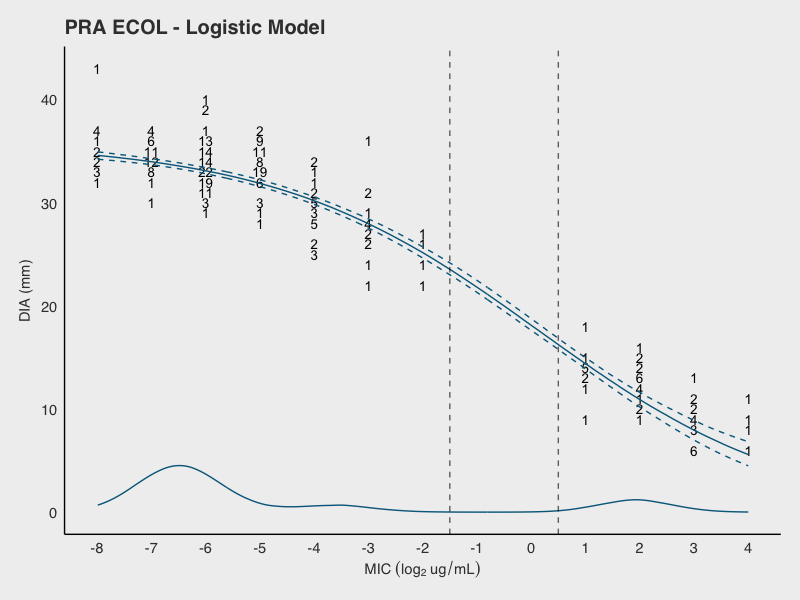} &   \includegraphics[scale=.25]{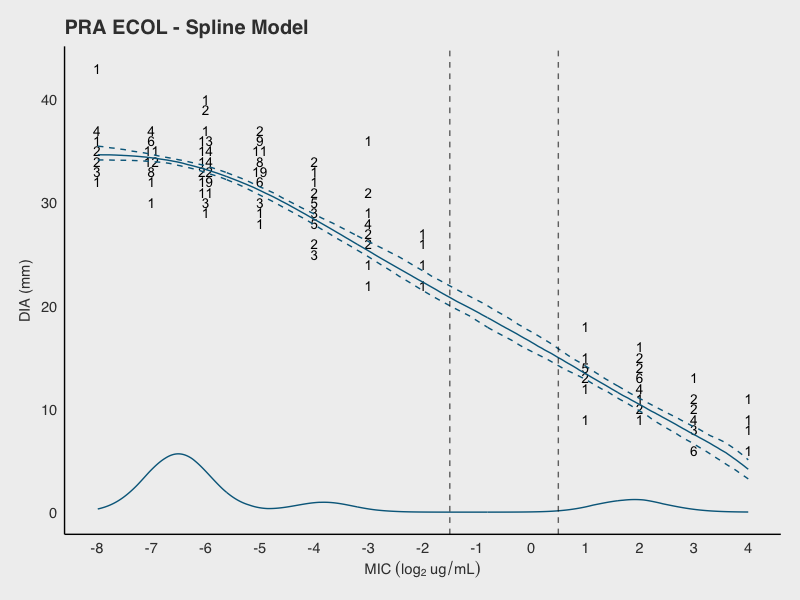} \\
\end{tabular}
\caption{\normalsize Model fits to the ERT EB (top) and PRA ECOL (bottom) data sets.  The logistic fit is on the left and the spline fit is one the right. For ERT EB, censored values are red colored (all are MIC censored, some of the 101 are DIA censored as well).  There is no censoring in the PRA ECOL data set.}
\label{fitModelPlot}
\end{centering}
\end{figure}

Figure \ref{fitModelPlot} shows the model fits to the two data sets.  The blue solid curve is the estimated true MIC-DIA relationship with the dashed lines representing 95\% credible intervals.  The estimated MIC density is plotted along the x-axis.  For both data sets, a large proportion of results are clearly in either the MIC susceptible or resistant regions.  Table \ref{modelFitTable} shows the posterior DIA breakpoint distributions.

\newpage

\begin{table}[ht!]
\captionsetup{justification=raggedright,singlelinecheck=false}
\centering
\begin{tabular}{l l r r | l r r}
\hline
&\multicolumn{3}{c}{\textit{Logistic}}&\multicolumn{3}{c}{\textit{Spline}} \\
& DIA & \% & Cumul. \%  &  DIA & \% & Cumul. \% \\
\hline
ERT EB & 19, 23 & 99.3 & 99.3 & 18, 21 & 51.0 & 51.0   \\
&&&& 18, 22 & 49.0 & 100  \\
\hline
\rowcolor{Gray}  PRA ECOL & 17, 25 & 46.4 & 46.4 & 16, 23 & 52.5 & 52.5  \\
\rowcolor{Gray}  & 17, 26 & 30.2 & 76.6 & 16, 22 & 28.1 & 80.6 \\
\rowcolor{Gray}  & 18, 26 & 15.3 & 91.8 & 17, 23 & 11.3 & 91.8 \\
\rowcolor{Gray}  & 18, 25 & 8.2 & 100.0 & 15, 22 & 3.2 & 95.0 \\
\bottomrule
\end{tabular}
\caption{DIA breakpoints (posterior top 95\%) for the two models.}
\label{modelFitTable}
\end{table}

The challenging feature of the ERT EB data set is the censoring at MIC values of -4 and 3.    However, because this censoring is far away from the MIC breakpoints, it should not impact the spline model inference.  Both models produce similar true MIC densities, although the spline model spikes higher and faster at the resistant endpoint due to not taking censoring into account.  The logistic model suggests an almost linear true relationship while the spline model suggests a more wavy fit.  This may indicate the data do not follow a logistic relationship.  The logistic model selected the set $(19,23)$ while the spline model selected the sets $(18, 21)$ and $(18,22)$ with roughly equal probability.  Given the indication the data do not follow a logistic relationship, the spline model breakpoints are recommended.  These breakpoints turn out to also be more closely aligned to the CLSI breakpoints of $(18, 20)$.

For the PRA ECOL data set, there is an absence of data around the MIC breakpoints -1 and 1. Given this gap in the MIC values, especially near the MIC breakpoints, our simulations suggest the logistic model should be preferred, even though both model fits in this case appear reasonable.   The recommended logistic breakpoints of $(17, 25)$ or $(17, 26)$ are much wider than the breakpoints  selected by CLSI.

There are several reasons why our recommended DIA breakpoints for these two data sets do not match the breakpoints selected by CLSI.  First, CLSI relied on the ERB method, which is known to produce inaccurate breakpoints \cite{craig99}. Second, CLSI does not distinguish between true and clinical MIC breakpoints. Finally, CLSI considers other factors, such as using breakpoints similar to other drugs or limiting the range on the indeterminant region. Although these model-based approaches do not consider these factors, they should provide initial breakpoint estimates more trustworthy than ERB when initiating discussions.  

\section{Conclusions}

Previous simulation studies by Craig and Qi have shown model-based procedures to be superior in both accuracy and precision to the commonly used error-rate bounded method.  In this paper, we expand these models to provide increased flexibility and therefore better accuracy.  Our proposed four-parameter logistic allows for asymmetric relationships that the three-parameter logistic did not.  Our Bayesian approach to jointly fit a monotonically decreasing spline and underlying MIC distribution avoids the possible loss of information inherent with the two-step approach.

We propose two ways to handle knot selection for the spline model.  The first uses many interior knots and uses a random walk prior with a smoothing parameter to avoid overfitting.  The second treats the number of interior knots and their locations as additional model parameters.  For the latter, a reversible jump algorithm is implemented.  This algorithm is complicated by the fact that the underlying data change each interaction of the Markov chain.

Both models are implemented in a freely available software package described in DePalma et al. \cite{DePalma16}.  This package provides the tools for clinicians to use the models in practice and compare to the ERB method.  All graphs produced in the real data analysis section of this paper are modified versions of those from this software.  

Currently, we are investigating methods to handle the spline fitting when there are censored observations.  While censoring can be ignored in most situations,  it will be a problem when there are a large number of censored observations and the MIC breakpoints are near an end of the data range (i.e., close to the censoring).  One approach is to convince researchers of the need to avoid censoring and to choose concentration ranges that minimize it.  As far as analytic methods, a hybrid model that combines parametric and nonparametric models may prove useful.

We are also researching more ways to evaluate the model fits to the data.  The Deviance Information Criterion (DIC) statistic is a widely used Bayesian information criterion for model selection.  With two competing models, the one with the smaller DIC statistic is usually preferred.  However, the logistic model incorporates censoring while the spline model does not.  Therefore, comparing the two in censoring situations is usually misleading.  It is also unclear how the DIC is related to precise breakpoint determination.  Our simulations have shown a good fit in the region of the MIC breakpoints is more important than a good overall fit.  

\newpage
\appendix
\section{Model Estimation}
\label{Appen: Estimation}
Because the posterior cannot be written in closed form, we approximate the posterior using Markov chain Monte Carlo (MCMC) \cite{liang10}.  Throughout the chain, a new estimate is proposed and either accepted or rejected.  To do this, we use Metropolis-Hastings updates \cite{HMCMC,hastings70}.

In order to construct our Markov chain, the first step is to initialize the unknown model parameters.  We then begin updating the parameters.  When possible we consider symmetric proposal distributions, which cancel out in the Metropolis-Hasting update procedure.  Here are the details of our MCMC processes, in terms of acceptance probabilities A, for the spline model (the logistic model is very similar):
\begin{itemize}
\item[1. ]  Update $\boldsymbol{m}$ (and  $\boldsymbol{d}$)  
\begin{itemize}
\item Proposal distribution: $q(m^*_i|m_i) = \mathcal{N}(m_i,.5)$
\end{itemize}
\begin{displaymath}
A=
\frac{\pi(m^*_i|f(\boldsymbol{m}))
\pi(\boldsymbol{x},\boldsymbol{y}|\boldsymbol{\beta},m^*_i)}
{\pi(m_i|f(\boldsymbol{m}))
\pi(\boldsymbol{x},\boldsymbol{y}|\boldsymbol{\beta},m_i)}
\end{displaymath}
\item[2. ] Update $\boldsymbol{\beta}$ 
\begin{itemize}
\item For Model 1, $\boldsymbol{\beta}$ is restricted to be positive and is computed on the log scale.  For Model 2, there is no restriction on $\boldsymbol{\beta}$ except the resulting fit is monotonic.
\item Proposal distribution - Model 1: $q(\boldsymbol{\beta^*}|\boldsymbol{\beta})= \mathcal{MN}(\boldsymbol{\beta},\Sigma)$ 
\item  Proposal distribution - Model 2: $q(\boldsymbol{\beta^*}|\boldsymbol{\beta})= \mathcal{MN}(\boldsymbol{\beta},\Sigma)$
\item Update Model 1:
\begin{displaymath}
A=
\frac{\pi(\boldsymbol{\beta^*,\lambda})
\pi(\boldsymbol{x},\boldsymbol{y}|\boldsymbol{\beta^*},\boldsymbol{m})}
{\pi(\boldsymbol{\beta,\lambda})
\pi(\boldsymbol{x},\boldsymbol{y}|\boldsymbol{\beta},\boldsymbol{m})}
\end{displaymath} 
\item Update Model 2:
\begin{displaymath}
A=
\frac{\pi(\boldsymbol{\beta^*})
\pi(\boldsymbol{x},\boldsymbol{y}|\boldsymbol{\beta^*},\boldsymbol{m})}
{\pi(\boldsymbol{\beta})
\pi(\boldsymbol{x},\boldsymbol{y}|\boldsymbol{\beta},\boldsymbol{m})}
\end{displaymath}
\item We consider an adaptive M-H step here where
\begin{equation*}
  \Sigma=
\begin{cases}
  \text{Diagonal covariance with variance equal to } 0.2 & i<1000 \\
  Cov(\boldsymbol{\beta})  \text{ of the past 500 estimates} & i \ge 1000
\end{cases}
\end{equation*}
\item If model 2 check to make that resulting fit is monotonically decreasing
\end{itemize}

\item[3. ] Update $\lambda$ (Model 1)
\begin{itemize}
\item Proposal distribution: $q(\lambda^*|\lambda) = \mathcal{U}(\lambda-.1,\lambda+.1)$
\begin{displaymath}
A=
\frac{\pi(\lambda^*)\pi(\boldsymbol{\beta}|\lambda^*)}
{\pi(\lambda)\pi(\boldsymbol{\beta}|\lambda)}
\end{displaymath}
\item Note: automatic reject if proposal is negative
\end{itemize}

\item[3. ] Update knot sequence $\boldsymbol{t}$ (Model 2)
\begin{itemize}
\item Described in Section \ref{sec:RJ}
\end{itemize}
\item[4. ] Update $f(\boldsymbol{m})$ - Dirichlet Process
\end{itemize}

\newpage
\section{Reversible Jump Step for Spline Knot Selection}
\label{sec:RJ}

Our RJMCMC procedure is based on the idea that the current spline coefficients, for given $\boldsymbol{m}$ and $\boldsymbol{y}$, should be relatively close to the LS coefficient estimates. We define $(t_1,t_2,\dots,t_{k})$ as the current interior knot sequence with $t_0$ and $t_{k+1}$ equal to the exterior knot locations. $\boldsymbol{\beta_{LS}}$ the current LS spline coefficients, $\boldsymbol{\beta}$ as the current spline coefficient estimates, and  $\Sigma_{LS}$ as the LS covariance matrix.  The current estimated true
DIA values are $\boldsymbol{d}$.

Once a new knot sequence has been proposed, we define $\boldsymbol{\beta^*_{LS}}$ as the new LS coefficients with corresponding covariance matrix $\Sigma^*_{LS}$, and the proposed new coefficients as  $\boldsymbol{\beta^{*}}$. $\boldsymbol{\beta^{*}}$ is drawn from a multivariate Normal distribution with $\boldsymbol{\mu}=\boldsymbol{\beta^*_{LS}}$ and $\Sigma=\Sigma^*_{LS}$.   The proposed new DIA values, based on $\boldsymbol{\beta^{*}}$, are $\boldsymbol{d^*}$.

\subsection{Birth}

We first consider the case of adding an interior knot to the existing knot sequence.  Given $t=(t_1,...,t_{k})$ a birth move proposes a new knot at location $t^*$, selected uniformly between $t_0$ to $t_{k+1}$. Assume the proposed $t^*$ falls between knots $t_j$ and $t_{j+1}$. The acceptance probability, $A$, is:

\begin{displaymath}
A=\frac{\pi(\boldsymbol{y}|\boldsymbol{d^*})\pi(\boldsymbol{d^*}|\boldsymbol{m},\boldsymbol{\beta^*})}
   {\pi(\boldsymbol{y}|\boldsymbol{d})\pi(\boldsymbol{d}|\boldsymbol{m},\boldsymbol{\beta})} \times
   \frac{\pi(\boldsymbol{\beta^{*}})\pi(k+1)}
   {\pi(\boldsymbol{\beta})\pi(k)}\times
   \frac{\mathcal{MN}(\boldsymbol{\beta}|\boldsymbol{\beta_{LS}},\Sigma_{LS})}
   {\mathcal{MN}(\boldsymbol{\beta^*}|\boldsymbol{\beta^*_{LS}},\Sigma^*_{LS})}
   \frac{\frac{1}{k+1}}
   {\frac{1}{k+1}\frac{1}{t_{j+1}-t_{j}}}\times
   |1|
\end{displaymath}

\subsection{Move}

A relocation move keeps the same number of interior knots, $k$, but the location of one knot is moved.  From the set of existing knots, an interior knot is randomly selected to be moved.  Denoting the location of the selected knot as $t_{j+1}$, the new knot location is proposed uniformly from $t_j$ to $t_{j+2}$.  The acceptance probability, $A$, is:

\begin{displaymath}
A=\frac{\pi(\boldsymbol{y}|\boldsymbol{d^*})\pi(\boldsymbol{d^*}|\boldsymbol{m},\boldsymbol{\beta^*})}
   {\pi(\boldsymbol{y}|\boldsymbol{d})\pi(\boldsymbol{d}|\boldsymbol{m},\boldsymbol{\beta})} \times
   \frac{\pi(\boldsymbol{\beta^{*}})}
   {\pi(\boldsymbol{\beta})}\times
   |1|
\end{displaymath}

\subsection{Death}

A death move removes one of the $k$ interior knots from the model.  Let $t_{j+1}$ be the location of the randomly selected knot to remove.  The acceptance probability, $A$, is:

\begin{displaymath}
A=\frac{\pi(\boldsymbol{y}|\boldsymbol{d^*})\pi(\boldsymbol{d^*}|\boldsymbol{m},\boldsymbol{\beta^*})}
   {\pi(\boldsymbol{y}|\boldsymbol{d})\pi(\boldsymbol{d}|\boldsymbol{m},\boldsymbol{\beta})} \times
   \frac{\pi(\boldsymbol{\beta^{*}})\pi(k-1)}
   {\pi(\boldsymbol{\beta})\pi(k)}\times
   \frac{\mathcal{MN}(\boldsymbol{\beta}|\boldsymbol{\beta_{LS}},\Sigma_{LS})}
   {\mathcal{MN}(\boldsymbol{\beta^*}|\boldsymbol{\beta^*_{LS}},\Sigma^*_{LS})}
   \frac{\frac{1}{k}\frac{1}{t_{j+1}-t_{j-1}}}{\frac{1}{k}}\times
   |1|
\end{displaymath}

\newpage
\bibliographystyle{WileyNJD-AMA}
\bibliography{BayesianMonotonicMeasurementError}

\end{document}